\documentclass[12pt]{article}

\pdfoutput=1

\usepackage{amsmath}
\usepackage{amsfonts, mathrsfs,amssymb}
\usepackage[dvips]{graphicx}
\usepackage{bbm}
\usepackage[authoryear,round]{natbib}
\usepackage[linkcolor=blue,citecolor=blue, colorlinks]{hyperref}
\usepackage{fancyref}
\date{}

\renewcommand{\ref}{\par\noindent\hangindent 12pt}

\newcommand{\noi}{\noindent}

\newcommand{\bfbbmY}{\mbox{\boldmath $\mathbbm{Y}$}}

\begin{document}
\title{A singular stochastic control approach for optimal 
pairs trading with proportional transaction costs}

\author{Haipeng Xing}

\maketitle

\centerline{Department of Applied Mathematics and Statistics}

\centerline{State University of New York, Stony Brook, NY 11794, USA.}

\centerline{xing@ams.sunysb.edu}

\bigskip
\noi {\bf Abstract}: 
Optimal trading strategies for pairs trading have been 
studied by models that try to find either optimal shares 
of stocks by assuming no transaction costs or 
optimal timing of trading fixed numbers of shares of 
stocks with transaction costs. 
To find optimal strategies which determine optimally 
both trade times and number of shares in pairs trading
process, we use a singular stochastic control approach 
to study an optimal pairs trading problem with 
proportional transaction costs. 
Assuming a cointegrated relationship for a pair of 
stock log-prices, we consider a portfolio optimization 
problem which involves dynamic trading strategies with 
proportional transaction costs. We show that the value 
function of the control problem is the unique viscosity 
solution of a nonlinear quasi-variational inequality, 
which is equivalent to a free boundary problem for the 
singular stochastic control value function. We then develop
a discrete time dynamic programming algorithm to compute 
the transaction regions, and show the convergence of the 
discretization scheme. We illustrate our approach with 
numerical examples and discuss the impact of different 
parameters on transaction regions. We study the out-of-sample
performance in an empirical study that consists of six 
pairs of U.S. stocks selected from different industry 
sectors, and demonstrate the efficiency of the optimal strategy.

\bigskip
\noi {\bf Keywords:} Free-boundary problem, pairs trading, stochastic control, 
trading strategies, transaction costs, transaction regions.

\newpage

\section{Introduction}

Pairs trading is one of proprietary statistical arbitrage 
tools used by many hedge funds and investment banks. It is 
a short-term trading strategy that first identifies two 
stocks whose prices are associated in a long-run equilibrium 
and then trades on temporary deviations of stock prices 
from the equilibrium. Though paris trading is a simple 
market neutral strategy, it has been used and discussed 
extensively by industrial practitioners in the last several 
decades; see detailed discussion in \cite{Vidyamurthy2004}, 
\cite{Whistler2004}, \cite{Ehrman2006}, 
\cite{laixing2008}, and reference therein. 

Besides its wide practice in financial industry, pairs trading 
also draws much attention from academic researchers. For 
instance, \cite{GatevEtAl2006} examined the risk and returns 
of pairs trading using daily data collected from the U.S. 
equity market and concluded that the strategy in general 
produces profit higher than transaction costs. To investigate 
the pairs trading strategy analytically, \cite{ElliottEtAl2005}
modeled the spread of returns as a mean-reverting process
and proposed a trading strategy based on the model. This motivates 
subsequent researchers to formulate pairs trading rules as 
stochastic control problems for an Ornstein-Uhlenbeck (OU) process 
and a correlated stock price process. In particular, 
\cite{MudchanatongsukEtAl2008} assumed the log-relationship 
between a pair of stock prices follows a mean-reverting process, 
and considered a self-financing portfolio strategy that only allows
positions that were long in one stock and short in the other with
equal dollar amounts. They then formulated a portfolio optimization 
based stochastic control problem and obtained the optimal solution 
to this control problem in closed form via the corresponding 
Hamilton-Jacobi-Bellman (HJB) equation. Relaxing the equal dollar
constraint, \cite{TourinYan2013} extended \cite{MudchanatongsukEtAl2008}'s
approach and study pairs trading strategies with arbitrary 
amounts in each stock without any transaction costs. 

Instead of deriving the optimal weight of stocks in pairs 
trading, another line of study on pairs trading strategies
fixes the number of traded shares for each stock during 
the entire trading process and considers only the optimal timing
of trades in the presence of fixed or proportional transaction
costs. Specifically, \cite{LeungLi2015} studies the optimal 
timing to open or close the position subject to fixed transaction 
costs and the effect of stop-loss level under the OU process 
by constructing the value function directly.
\cite{ZhangZhang2008}, \cite{SongZhang2013}, and \cite{NgoPham2016}
studied the optimal pairs trading rule that is based on optimal switching
among two (buy and sell) or three (buy, sell, and flat) regimes with 
a fixed commission cost for each transaction, and solve the problem
by finding viscosity solutions to the associated HJB equations
(quasi-variational inequalities). \cite{LeiXu2015} studied the 
optimal pairs trading rule of entering and exiting the asset 
market in finite horizon with proportional transaction cost 
for two convergent assets. Note that, although transaction 
costs are considered in these strategies, since the number of 
traded shares of stocks are fixed during the entire trading
period, these strategies are still far from traders' 
practical experience in reality.

To bridge the gap between choosing optimal weight of shares 
and deciding optimal trading times in pairs trading, we use 
a singular stochastic control approach to study 
an optimal pairs trading problem with proportional transaction 
costs which allows us choosing not only optimal weight, but also
optimal trading times during the trading process. 
For convenience, we assume the same diffusion and 
Urnstein-Uhlenbeck processes for one stock and its 
spread with the other stock as those in 
\cite{MudchanatongsukEtAl2008}. However, different from
\cite{MudchanatongsukEtAl2008} who used a trading rule
which requires to short one stock and long the other 
in equal dollar amounts, we consider a delta-neutral
rule under which the ratio of traded shares for two
stocks is fixed and this fixed ratio is determined by 
the cointegration relationship of two stocks. 
Hence when the number of shares of one stock is determined,
based on the rule of delta neutral, the number of shares 
for the other stock is also determined. 
Besides the weight of shares need to be optimally chosen,
we also assume a proportional transaction cost for each 
trade and hence the optimal times of trading also needs to 
be decided. 

As the overall transaction cost based on the above assumption
depends on both trading times and the numbers of shares in 
each trade, we compute the terminal utility of wealth over 
a fixed horizon and formulate the problem of choosing optimal 
trading times and the number of shares as a singular stochastic
control problem. 
We derive the Hamilton-Jacobi-Bellman equations for this 
problem, and show that the value function of the problem is 
the unique viscosity solution of a quasi-variational inequality. 
We further argue that the quasi-variational inequality is 
equivalent to a free boundary problem so that the state space 
consisting of one stock price and its spread with the other 
stock can be naturally divided into three transaction regions: 
long the first stock and short the second, short the first 
and long the second, and no transaction. 
The implied transaction regions can help us determine not only 
optimal times of each transaction, but also the optimal number of shares
in each transaction. To compute the boundaries of these 
transaction regions, we develop a numerical algorithm that 
is based on discrete time dynamic programming to solve the 
equation for the negative exponential utility function,
and show that the numerical solution converges to the unique 
continuous-time solution of the problem.

To investigate the performance of the optimal trading strategies implied
by the transaction regions, we carry out both simulation and empirical
studies. Specifically, we study the time-varying transaction regions 
(or trading boundaries) for a specific set of model parameters, and 
investigate the impact of variations of model parameters on transaction 
regions and performance of the optimal strategy. For comparison purpose, 
we also consider a benchmark strategy which is based on the deviation of
the spread from its long-term mean and is popular among practitioners. 
In both simulation studies and real data analysis, we show that the optimal
trading strategy performs better than the benchmark strategy.

The rest of the paper is organized as follows. Section 2 first 
formulates the model and then derive the Hamilton-Jacobi-Bellman 
equations associated with the singular stochastic control problems. 
It shows the existence and uniqueness of the viscosity solution for 
the varational inequalities which are equvalent to the portfolio 
optimization problem, and reduces the problem into a free boundary 
problem. Section 2 also consider the optimal trading problem 
with exponential utility functions. In section 3, we discretize 
the free boundary problem and propose a disctete time 
dynamic programming algorithm. We also demonstrate that
the solution of the discretized problem converges to the
viscosity solution of the variational inequalities. Sections 4
and 5 provide simulation and empircal studies of the model 
and the optimal trading strategy, and compare its performance
with a benchmark trading strategy.
Some concluding remarks are given in Section 6.

\section{A pairs trading problem with proportional
transaction costs}

\subsection{Model specification}

Consider a pair of two stocks $P$ and $Q$, and let $p(t)$ and $q(t)$ 
denote their prices at time $t$, respectively. We assume that the price of 
stock $P$ follows a geometric Brownian motion,
\begin{equation}\label{stock1.dyn}
dp(t) = \mu p(t) dt + \sigma p(t) dB(t),
\end{equation}

\noi where $\mu$ and $\sigma$ are the drift and the volatility of
stock $P$, and $B(t)$ is a standard Brownian motion defined on a
filtered probability space and but specified later. Denote 
$x(t)$ the difference of the logarithms of the two stock prices, i.e., 
\begin{equation}\label{spread.def}
x(t) = \log q(t) - \log p(t) = \log (q(t)/p(t)).
\end{equation}

\noi We assume that the spread follows an Ornstein-Uhlenbeck process
\begin{equation}\label{spread.dyn}
dx(t) = \kappa (\theta-x(t)) dt + \nu dW(t),
\end{equation}

\noi where $\kappa>0$ is the speed of mean reversion, and $\theta$
is the long-term equilibrium level to which the spread reverts.
We assume that $(B(t), W(t))$
is a two-dimensional Brownian motion defined on a filtered probability
space $(\Omega, {\cal F}_t, \mathbbm{P})$, and the instantaneous 
correlation coefficient between $B(t)$ and $W(t)$ is $\rho$, i.e., 
\begin{equation}
E[ dW(t) dB(t) ] = \rho dt.
\end{equation}

\noi The above assumptions are same as those in 
\cite{MudchanatongsukEtAl2008}. With these assumptions, we can 
express the dynamics of $q(t)$ as
\begin{equation}\label{stock2.dyn}
dq(t) = \big[ \mu + \kappa (\theta-x(t)) + \frac{1}{2} \nu^2 +
\rho \sigma \nu \big] q(t) dt + \sigma q(t) dB(t) + \nu q(t) dW(t).
\end{equation}

In the presence of proportional transaction costs, the investor pays 
$0<\zeta_p, \zeta_q<1$ and $0<\eta_p, \eta_q<1$ of the dollar value transacted 
on purchase and sale of the underlying stocks $P$ and $Q$.
Denote $L_p(t)$ and $M_p(t)$  two nondecreasing and non-anticipating 
processes and represent the cumulative number of shares of stock 
$P$ bought or sold, respectively, within the time interval $[0, t]$, 
$0\le t \le T$. Let $y_p(t)$ be the number of shares held in stock $P$, 
i.e., $y_p(t) = L_p(t) - M_p(t)$, and similarly, we define $L_q(t)$, 
$M_q(t)$, and $y_q(t) = L_q(t)-M_q(t)$ for stock $Q$. Denote 
$g(t)$ the dollar value of 
the investment in bond which pays a fixed risk-free rate of $r$. 
Then the investor's position in two stocks and the bond is driven by
\begin{equation}\label{stock.buy.sell}
dy_p(t) = dL_p(t) - d M_p(t), \qquad dy_q(t) = dL_q(t) - d M_q(t)
\end{equation}

\noi and 
\begin{equation}\label{bond.pos1}
dg(t) = rg(t)dt + b_p p(t) dM_p(t) - a_q q(t) dL_q(t) 
+b_q q(t) dM_q(t) - a_p p(t) dL_p(t), 
\end{equation}

\noi where $a_i=1+\zeta_i$ and $b_i=1-\eta_i$ for $i=p, q$. 

We then need to choose a rule to determine the number of shares of stocks 
$P$ and $Q$ bought or sold at time $t$. Note that, \cite{MudchanatongsukEtAl2008} assumed no transaction cost and considered the strategy that always shorts 
one stock and longs the other in equal dollar amount, i.e., 
$p(t) dL_p(t)+q(t) dM_q(t)=0$ or $p(t) dM_p(t) + q(t) dL_q(t)=0$
at time $t$. \cite{LeiXu2015} and \cite{NgoPham2016} considered
a delta-neutral strategy that always long one share of a stock and
short one share of the other stock, i.e., 
$dy_p(t)= - dy_q(t)=1$ or $dy_p(t)= - dy_q(t)=-1$ at time $t$. 
Here, we also consider a delta-neutral strategy that requires the total 
of positive and negative delta of two assets is zero, hence it suggests
that the number of shares of stock $P$ bought (or sold) 
at time $t$ are same as the number of shares of stock $Q$ sold 
(or bought), i.e., 
\begin{equation}\label{delta.neutral}
d L_p(t) = d M_q(t), \qquad 
d M_p(t) = d L_q(t). 
\end{equation}

\noi Equation \eqref{delta.neutral} implies that 
$$dy_q(t) = -dy_p(t) $$ 

\noi at any time $t$. Comparing to \cite{LeiXu2015} and 
\cite{NgoPham2016}, we remove the constraint $dy_p(t)= - dy_q(t)=1$
or $-1$ and allow $y_p(t)= - y_q(t)$ to be a control variable. 
Using equations \eqref{stock2.dyn}
and \eqref{delta.neutral}, the dynamics of $g(t)$ in equation 
\eqref{bond.pos1} can be simplified as 
\begin{equation}\label{bond.pos2}
dg(t) = rg(t) dt - \big( a_p - b_q e^{x(t)} \big) p(t)
dL_p(t) + \big( b_p - a_q e^{x(t)} \big) p(t) dM_p(t).
\end{equation}

\noi The process $(L_p(t), M_p(t))$ together with our delta-neutral
strategy provides us an admissible trading strategy. For convenience, we denote 
${\cal T}(g_0)$ the set of admissiable trading strategies that an 
investor starts at time zero with amount $g_0$ of the investment 
in bond and zero holdings in two stocks (i.e., $y_p(0)=y_q(0)=0$), which
indicates that the numbers of shares held in stocks $P$ and $Q$ at time 
$t$ are $y_p(t)$ and $-y_p(t)$, respectively. For nonational convenience,
we omit  the subscript of $y_p(t)$ and denote $y_p(t)$ as $y(t)$ in 
our discussion. Then equations \eqref{stock1.dyn},
\eqref{spread.dyn}, \eqref{stock.buy.sell}, and \eqref{bond.pos2}
compose the market model in the time interval $[0, T]$, which describes
a stochastic process of $(p(t), x(t), y_p(t), g(t))$ in $\mathbbm{R}^+
\times \mathbbm{R} \times \mathbbm{R} \times \mathbbm{R}$. 

Denote the terminal value of the pairs trading portfolio 
by $J(x(T), p(T), y(T))$. Note that, under our assumption, 
$y(T)$  indicates that the investor's positions in stocks $P$ and $Q$ 
are $y(T)$ and $-y(T)$, respectively,
then the liquidated value of the portfolio is
\begin{equation}
J(p(T), x(T), y(T)) = A_+ (p(T), x(T)) y(T) \mathbbm{1}_{\{
y(T) \ge 0 \}} + A_- (p(T), x(T)) y(T) \mathbbm{1}_{\{
y(T) < 0 \}},
\end{equation}

\noi where 
$$
A_+ (p, x) = ( b_p - a_q e^{x}) p, \qquad
A_- (p, x) = ( a_p - b_q e^{x}) p.
$$

\noi Furthermore, if the investment in bond at terminal time $T$ is 
$g(T)$, the terminal wealth of the investor is given by 
$g(T) + J(p(T), x(T), y(T))$.
Suppose that the investor's utility $U: \mathbbm{R} \longrightarrow 
\mathbbm{R}$ is a concave and increasing function with $U(0)=0$. 
We assume that the investor's goal is to maximize the expected utility
of terminal wealth under the market model \eqref{stock1.dyn},
\eqref{spread.dyn}, \eqref{stock.buy.sell}, and \eqref{bond.pos2}, 
\begin{equation}\label{opt.equ}
\begin{aligned}
V(t, p, x, y, g) &= \sup_{(L_p(t), M_p(t)) \in {\cal T}(g_0)} E 
\Big\{ U( g(T) + J(p(T), x(T), y(T)) | p(t)=p, \\
& \hspace{40pt} x(t)=x, y_t = y, g(t)= g \Big\}.
\end{aligned}
\end{equation}

\noi Furthermore, given  trading strategies $(L_p, M_p)$, the total trading cost incurred 
over $[t, T]$ can be expressed as
\begin{eqnarray}\label{totalcost}
C(L_p, M_p; t, T) &=& \int_t^T e^{r(T-u)} A_-(p(u), x(u)) dL_p(u) 
- \int_t^T e^{r(T-u)} A_+(p(u), x(u)) dM_p(u) \nonumber \\ 
&& \hspace{30pt} - J(p(T), x(T), y(T)).
\end{eqnarray}

\noi and the total profit over $[t, T]$ is $-C(L_p, M_p; t, T)$.

\subsection{The Hamilton-Jacobi-Bellman equations and free boundary problems}

We now derive the Hamilton-Jacobi-Bellman (HJB) equations, associated with 
the stochastic control problems, for the utility maximization problem 
\eqref{opt.equ}. Consider a class of trading strategies such that 
$L_p(t)$ and $M_p(t)$ are absolutely continuous processes, given by
$$
L_p(t) = \int_0^t l(u) du, \qquad M_p(t) = \int_0^t m(u) du,
$$

\noi where $l(u)$ and $m(u)$ are positive and uniformly bounded by 
$\xi< \infty$. Then \eqref{stock1.dyn},
\eqref{spread.dyn}, \eqref{stock.buy.sell}, and \eqref{bond.pos2}
provides us a system of stochastic differential equations with controlled
drift, and the Bellman equation for a value function denoted by $V^{\xi}$
is 
\begin{equation*}
{\cal L}_{1,o} V^\xi + \sup_{0\le l_t, m_t \le \xi} \Big\{ \Big[ {\cal L}_{1,b} 
V^\xi \Big] l_t - \Big[ {\cal L}_{1,s} V^\xi \Big] m_t \Big\} =0,
\end{equation*}

\noi for $(t, p, X, y, g) \in [0, T] \times \mathbbm{R}^+ \times
\mathbbm{R} \times \mathbbm{R} \times \mathbbm{R}$, 
in which the operators ${\cal L}$, ${\cal B}$, and ${\cal S}$ are defined as
\begin{equation*}\label{operA.def}
{\cal L}_{1,o} := 
\frac{\partial }{\partial t} + \kappa \big( \theta-x \big) \frac{\partial
  }{\partial x} + \mu p \frac{\partial }{\partial p} + 
rg \frac{\partial }{\partial g} + \frac{1}{2} \nu^2
\frac{\partial^2 }{\partial x^2} + \rho \nu \sigma p
\frac{\partial^2 }{\partial p \partial x} + \frac{1}{2} 
\sigma^2 p^2 \frac{\partial^2 }{\partial p^2},
\end{equation*}
$$
{\cal L}_{1,b} := \frac{\partial }{\partial y} - \big( a_p - b_q e^{x(t)} \big) p(t) \frac{\partial }{\partial g},
$$
$$
{\cal L}_{1,s} :=\frac{\partial }{\partial y} - \big( b_p - a_q e^{x(t)} \big) p(t) \frac{\partial }{\partial g}.
$$

\noi The optimal trading strategy is then determined by considering 
the following three possible cases: 
\begin{enumerate}
\item [(i)] buying stock $P$ and sell stock $Q$ at the same rate
$l(t)=\xi$ (i.e., $m(t)=0$) when 
\begin{equation}\label{reg1.equ}
{\cal L}_{1,b} V^\xi \ge 0, \qquad {\cal L}_{1,s} V^\xi > 0;
\end{equation}

\item [(ii)] selling stock $P$ and buy stock $Q$ at rate 
$m(t)=\xi$ (i.e., $l(t)=0$) when
\begin{equation}\label{reg2.equ}
{\cal L}_{1,b} V^\xi < 0, \qquad {\cal L}_{1,s} V^\xi \le  0;
\end{equation}

\item [(iii)] doing nothing (i.e. $l(t)=m(t)=0$) when 
\begin{equation}\label{reg3.equ}
{\cal L}_{1,b} V^\xi \le 0, \qquad {\cal L}_{1,s} V^\xi \ge 0.
\end{equation}
\end{enumerate}

\noi Note that the case ${\cal L}_{1,b} V^\xi >0$ and ${\cal L}_{1,s} V^\xi
<0$ can not occur, as all value functions are increasing 
functions of $g$. 

The above argument shows that the optimization problem 
\eqref{opt.equ} is a free boundary problem in which the 
optimal trading strategy is defined by the inequalities
(i), (ii), and (iii) for a given value function. 
Besides, the state space 
$[0, T] \times \mathbbm{R}^+ \times
\mathbbm{R} \times \mathbbm{R} \times \mathbbm{R}$
is partitioned into {\it buy}, {\it sell}, and 
{\it no-transaction} regions for stock $P$, 
which are characterized by inequalities \eqref{reg1.equ}, 
\eqref{reg2.equ}, and \eqref{reg3.equ}, respectively. 
For sufficiently large $\xi$, the
state space remains divided into a {\it buy region} ${\cal B}$,
a {\it sell region} ${\cal S}$, and a {\it no-transaction
region} ${\cal N}$ for stock $P$, which are 
correspondingly the {\it sell region}, the {\it buy region},
and the {\it no transaction region} for stock $Q$ 
due to equation \eqref{delta.neutral}. Obviously, the 
buy and sell regions for stock $P$ are disjoint, as it is
not optimal to buy and sell the same stock at the same time. 
We denote the boundaries between the no-transaction region
${\cal N}$ and the buy and sell regions ${\cal B}$ and ${\cal S}$
as $\partial {\cal B}$ and $\partial {\cal S}$, respectively.

Let $\xi \rightarrow \infty$, the class of admissible trading strategies
becomes ${\cal T}(g_0)$. We can guess that the state space is still 
divided into three regions,  a region of buying $P$ and selling $Q$, 
a region of selling $P$ and buying $Q$, and a no-transaction region. 
Then the optimal trading strategy requires an immediate move to 
the boundaries of buy or sell regions, 
if the state is in the buy region ${\cal B}$ or the sell region ${\cal S}$. 
Actually we can obtain equations that each of the value functions should
satisfy as follows. 

(i) In region ${\cal B}$ of buying $P$ and selling $Q$, the value function
remains constant along the path of the state, dictated by the optimal
trading strategy, and therefore, for $\delta y \ge 0$
\begin{equation}\label{reg1.equ1}
V(t, p, x, y, g) = V(t, p, x, y+ \delta y, g - ( a_p - b_q e^{x} ) p \delta y),
\end{equation}

\noi where $\delta y$ is the number of shares of stock $P$ bought 
and stock $Q$ sold by the investor. $\delta y$ can be any positive value 
up to the number required to take the state to $\partial {\cal B}$, 
so letting $\delta y \rightarrow 0$ in \eqref{reg1.equ1} yields
\begin{equation}\label{reg1.equ2}
{\cal L}_{1,b} V = 0.
\end{equation}

(ii) Similarly, in region ${\cal S}$ of selling $P$ and buying $Q$, the value 
function obeys the following equation for $\delta y \ge 0$
\begin{equation}\label{reg2.equ1}
V(t, p, x, y, g) = V(t, p, x, y- \delta y, g + (b_p - a_q e^{x} ) p \delta y), 
\end{equation}

\noi where $\delta y$ is the number of shares of stock $P$ sold 
and stock $Q$ bought by the investor. $\delta y$ can be any positive 
value up to the numer required to take the state to $\partial {\cal S}$, so
letting $\delta y \rightarrow 0$ in \eqref{reg2.equ1} yields 
\begin{equation}\label{reg2.equ2}
{\cal L}_{1,s} V = 0.
\end{equation}

(iii) In the no-transaction region, the value function obeys the same
set of equations obtained for the class of absolutely continuous
trading strategies, and therefore the value function is given by
\begin{equation}\label{reg3.equ1}
{\cal L}_{1,o} V = 0,
\end{equation}

\noi and the pair of inequalities, shown above in \eqref{reg3.equ},
also hold. Note that, due to the continuity of the value function, if
it is known in the no-transaction region, it can be determined in both
the buy and sell regions by \eqref{reg1.equ2} and \eqref{reg2.equ2},
respectively. 

In the buy region ${\cal B}$, ${\cal L}_{1,s} V <0$, and, in the sell region
${\cal S}$, ${\cal L}_{1,b} V > 0$. Also, from the two pairs of
inequailities \eqref{reg1.equ} and \eqref{reg2.equ}, we may conjecture
that ${\cal L}_{1,o} V$ in  \eqref{reg3.equ1} is negative in both the
buy region ${\cal B}$ and the sell region ${\cal S}$. 
Therefore, the above set of equations can be
summarized as  the following fully nonlinear partially differential 
equations (PDE):
\begin{equation}\label{var.equ}
\min \Big\{ -{\cal L}_{1,b} V, {\cal L}_{1,s} V, -{\cal L}_{1,o} V \Big\} =0
\end{equation}

\noi for $(t, p, X, y, g) \in [0, T] \times \mathbbm{R}^+ \times
\mathbbm{R} \times \mathbbm{R} \times \mathbbm{R}$.
Note that the above discussion also yields the following free
boundary problem for the singular stochastic control value 
function:
\begin{equation}\label{free.bound1}
\left\{ \begin{array}{rcll}
{\cal L}_{1,b} V &=& 0 & \mbox{in } {\cal B} \\
{\cal L}_{1,s} V &=& 0 &  \mbox{in } {\cal S} \\
{\cal L}_{1,o} V &=& 0 &  \mbox{in } {\cal N} \\
V(T, p, x, y, g) &=& U( g + J(p, x, y)).
\end{array} \right.
\end{equation}

\bigskip
We next show that the value function given by \eqref{opt.equ} is
a constrained viscosity solution of the variational inequality 
\eqref{var.equ} on $[0, T] \times \mathbbm{R}^+ \times
\mathbbm{R} \times \mathbbm{R} \times \mathbbm{R}$, 
and it is the unique bounded constrained viscosity solution 
of \eqref{var.equ}. The proof is given in the appendix.

\medskip
{\bf Theorem 1}. The value function $V(t, p, x, y, g)$ is a
constrained viscosity solution of \eqref{var.equ}
on $[0, T] \times \mathbbm{R}^+ \times
\mathbbm{R} \times \mathbbm{R} \times \mathbbm{R}$.

\medskip
{\bf Theorem 2}. Let $u$ be a bounded upper semicontinuous viscosity
subsolution of \eqref{var.equ}, and $v$ a bounded from below lower
semicontinuous viscosity supersolution of \eqref{var.equ}, such that
$u(T, {\bf x}) \le v(T, {\bf x})$ for all ${\bf x} \in \mathbbm{R}^+ \times
\mathbbm{R} \times \mathbbm{R} \times \mathbbm{R}$. Then $u \le v$ on $[0, T] \times \mathbbm{R}^+ \times
\mathbbm{R} \times \mathbbm{R} \times \mathbbm{R}$.

\subsection{Optimal trading with exponential utility functions} 

We next assume that the investor has the negative exponential 
utility function 
\begin{equation}\label{cara.util}
U(z) = 1-\exp( - \gamma z),
\end{equation}

\noi where $\gamma$ is the constant absoluate risk aversion (CARA)
parameter such that $-U''(z)\big/U'(z) = \gamma$. For equation 
\eqref{var.equ}, this utility function can reduce much of
computational effort and is easy to interpret. Note that
for the utility function \eqref{cara.util}, the definition of
the value function \eqref{opt.equ} can be expressed as 
\begin{equation}\label{exputl.equ1}
V(t, p, x, y, g) = 
1- \exp \Big( -\gamma g e^{r(T-t)} \Big) H(t, p, x, y),
\end{equation}

\noi where $H(t, p, x, y)$ is a convex nonincreasing continuous
function in $y$ and defined by
\begin{eqnarray*}
H(t, p, x, y) &=& \inf_{Lp(t), M_p(t) \in {\cal T}(g_0)} 
E \Big\{ \exp[  - \gamma J(p(T), x(T), y(T)] \big|
p(t) = p, x(t)=x, y(t) = y \Big\} \nonumber \\
&=&1-V(t, p, x, y, 0). 
\end{eqnarray*}

\noi Plug \eqref{exputl.equ1} into \eqref{var.equ}, and
define the following operators for $H(t, p, x, y)$ on 
$[0, T] \times \mathbbm{R}^+ \times \mathbbm{R} \times
\mathbbm{R}$, 
$$
{\cal L}_{2,o} H = \frac{\partial H}{\partial t} + 
\kappa (\theta-x) \frac{\partial H}{\partial x} + \mu p
\frac{\partial H}{\partial p} + \frac{1}{2} \nu^2
\frac{\partial^2 H}{\partial x^2} + \rho \nu \sigma p 
\frac{\partial^2 H}{\partial p \partial x} + \frac{1}{2} 
\sigma^2 p^2 \frac{\partial^2 H}{\partial p^2},
$$
$$
{\cal L}_{2,b} H = \frac{\partial H}{\partial y} + 
\gamma e^{r(T-t)} A_-(p,x) H, 
$$
$$
{\cal L}_{2,s} H = \frac{\partial H}{\partial y} + 
\gamma e^{r(T-t)} A_+(p,x) H.
$$

\noi Then \eqref{var.equ} is transformed into the following 
PDE for $H(t, p, x, y)$
\begin{equation}\label{var.equ3}
\min \Big\{ {\cal L}_{2,b} H, -{\cal L}_{2,s} H, 
{\cal L}_{2,o} H \Big\} =0
\end{equation}

\noi with the following boundary conditions
$$
H(T, p, x, y) = \exp \big\{ -\gamma J(p, x, y) \big\}.
$$

\noi Correspondingly, the free boundary problem \eqref{free.bound1}
becomes
\begin{equation}\label{free.bound2}
\left\{ \begin{array}{rcll}
{\cal L}_{2,o} H &=& 0 &   y\in [ Y_b(t, p, x), Y_s(t, p, x)]  \\
{\cal L}_{2,b} H &=& 0 & y \le Y_b(t, p, x) \\
{\cal L}_{2,s} H &=& 0 & y \ge Y_s (t, p, x) \\
H(T, p, x, y) &=& \exp \big\{ -\gamma J(p, x, y) \big\}.
\end{array} \right.
\end{equation}

\noi in which $Y_b(t, p, x)$ and $Y_s(t, p, x)$ are 
the buy and sell boundaries for stock $P$, respectively. 
Note that the function $H(t, p, x, y)$ is evaluated in the
four-dimensional space $[0, T] \times \mathbbm{R} \times \mathbbm{R}
\times \mathbbm{R}$. Furthermore, this suggests that 
while $(t, u_t, w_t)$ is inside the no-transaction region, 
the dynamics of $h(t, u, w, y)$ is driven by a two-dimensional 
standard Brownian motions $\{ z_t, t \ge 0 \}$ and 
$\{ w_t, t\ge 0 \}$ with correlation $\rho$. In the buy and 
sell regions, it follows from \eqref{free.bound2} that 
\begin{eqnarray*}
&&H(t,p,x,y) =\exp \{ -\gamma e^{r(T-t)}A_-(p, x) [ y-Y_b(t,p,x)] \} 
H(t,p,x,Y_b(t, p, x)), \quad y \le Y_b(t, p, x),\\
&&H(t,p,x,y) = \exp \{ -\gamma e^{r(T-t)} A_+(p, x) [ y-Y_s(t,p,x)] \} 
H(t,p,x,Y_s(t, p, x)), \quad y \ge Y_s(t, p, x).
\end{eqnarray*}

\section{Discretization and a numerical algorithm}\label{num.alg}

The solution of the PDE \eqref{var.equ} or \eqref{var.equ3} can be obtained by 
turning the stochastic differential equations \eqref{stock1.dyn},
\eqref{spread.dyn}, \eqref{stock.buy.sell}, and \eqref{bond.pos2}
into Markov chains and then applying the discrete time dynamic programming algorithm. 
The discrete state is
$\mathbbm{X}=(\chi, \mathbbm{p}, \mathbbm{x},  \vartheta, \mathbbm{g})$,
whose elements denote time, price of stock $P$, spread, 
number of shares of stock $P$, and amount in the bank in 
a discrete space. The value function, denoted by $\mathbbm{V}$, 
are given a value at the final time by using
the boundary conditions for the continuous value functions over the
discrete subspace $(\mathbbm{p}, \mathbbm{x}, \vartheta, \mathbbm{g})$, and
then they are estimated by proceeding backward in time by using the
discrete time algorithm. As in the continuous time case, this
algorithm is the same for both value functions and is derived below
for a value function denoted by $\mathbbm{V}^{\delta}(\chi, \mathbbm{p},
\mathbbm{x},  \vartheta, \mathbbm{g})$, where $\rho$ is a discretization
parameter, which depends on the discrete time interval $t_\delta$. If
$t_\delta$ and the resolution of the $\vartheta$-axis $\vartheta_\delta$ are
sent to zero, then the above discrete value function converges to a
viscosity subsolution and a viscosity supersolution of the
PDE \eqref{var.equ}. Therefore, all the discrete value functions
converge to their continuous counterparts; this is due to the
uniqueness of the viscosity solution. 

Consider an evenly spaced partition of the time interval $[0, T]$: $\chi = \{ 
\delta,  2\delta, \dots, n\delta \}$, where $\delta = T/n$, and
two evenly spaced partitions of 
the space intervals $\mathbbm{z}= \{ 0, \pm \sqrt{\delta}, \pm 2 \sqrt{\delta},
\dots, \}$ and $\mathbbm{w} = \{ 0, \pm \sqrt{\delta}, \pm 2\sqrt{\delta},
\dots, \}$. The grids $\mathbbm{p}$ is defined by $\mathbbm{z}$ via
the following transformation,
\begin{equation}\label{dis.stock.price1}
\mathbbm{p}_i = \exp \Big( 
(\mu-\frac{1}{2} \sigma^2)  T+ \mathbbm{z}_i \sigma \sqrt{T} \Big).
\end{equation}

\noi Note that the SDE \eqref{spread.dyn} implies that the aymptotic 
distribution of $X(t)$ is Normal$(\theta, \nu^2/(2\kappa))$, we define 
grid $\mathbbm{x}$ by 
\begin{equation}\label{dis.spread}
\mathbbm{x}_j = \theta + \frac{\nu}{\sqrt{2\kappa}} \mathbbm{w}_j.
\end{equation}

\noi Denote $\chi_i = i\delta$ for $i=1, \dots, n-1$. The dynamics 
\eqref{stock1.dyn} and \eqref{spread.dyn} of $P(t)$ and $X(t)$ implies 
the following transition density for $(\mathbbm{p}(\chi_i), 
\mathbbm{x}({\chi_i}) )$, 
\begin{equation}\label{trans.den}
{ \mathbbm{p}({\chi_{i+1}}) \choose \mathbbm{x}_{\chi_{i+1}} } \Bigg|
{ \mathbbm{p}({\chi_i}) \choose \mathbbm{x}_{\chi_i} } \sim 
N \Bigg( { \log \mathbbm{p}(\chi_i) + (\mu-\frac{1}{2} \sigma^2) \delta
\choose (1-\delta \kappa) \mathbbm{x}({\chi_i}) + \delta \kappa \theta }, 
\left( \begin{array}{rr} \delta \sigma^2, \delta \rho \sigma \nu \\
 \delta \rho \sigma \nu, \delta \nu^2 \end{array} \right) \Bigg).
\end{equation}

\noi We also note that the discrete time equation for the amount in the 
bank $\mathbbm{g}(\chi)$ is 
$$
\mathbbm{g}(\chi_{i+1}) = \mathbbm{g}(\chi_i) \exp ( r \delta).
$$

Given the grid defined above, 
the discrete time dynamic programming principle is invoked, and the
following discretization scheme is proposed for PDE 
\eqref{var.equ}:
\begin{equation}\label{var.equ4}
\begin{aligned}
& \mathbbm{V}^\delta(\chi_i, \mathbbm{p}(\chi_i), \mathbbm{x}(\chi_i), 
\vartheta, \mathbbm{g}(\chi_i) ) = \max \Big\{ \\
& \hspace{30pt}  \mathbbm{V}^\delta
\big(\chi_i, \mathbbm{p}(\chi_i), \mathbbm{x}(\chi_i), 
\vartheta + \xi, \mathbbm{g}(\chi_i) - ( a_p-b_q 
e^{\mathbbm{x}(\chi_i)} ) \mathbbm{p}(\chi_i) \xi \big), \\ 
& \hspace{30pt}  \mathbbm{V}^\delta
\big(\chi_i, \mathbbm{p}(\chi_i), \mathbbm{x}(\chi_i), 
\vartheta-\xi, \mathbbm{g}(\chi_i) + ( b_p -a_q
e^{\mathbbm{x}(\chi_i)}) \mathbbm{p}(\chi_i) \xi \big), \\
& \hspace{30pt} 
E \big\{ \mathbbm{V}^\delta \big(\chi_{i+1}, \mathbbm{p}(\chi_{i+1}), 
\mathbbm{x}(\chi_{i+1}), \vartheta, \mathbbm{g}(\chi_{i+1}) \big) 
\big\} \ \Big\}.
\end{aligned}
\end{equation}

\noi where $\xi >0$ is a real constant and $i=0, \dots, n-1$. 
This scheme is based on the
principle that the investor's policy is the choice of the optimum
transaction. We next show that, as the discretization parameter
$\delta \rightarrow 0$, the solution $\mathbbm{V}^\delta$ of
\eqref{var.equ4} converges to the value function $V$, or,
equivalently, to the unique constrained viscosity solution of
\eqref{var.equ}. 

\medskip
{\bf Theorem 3}. The solution $\mathbbm{V}^\delta$ of
\eqref{var.equ4} converges locally uniformly as $\delta \rightarrow 0$
to the unique continuous constrained viscosity solution of
\eqref{var.equ}. 
\medskip

For the exponential utility function $U(z) = 1-\exp( -\gamma z)$, the
value function $V$ can be expressed as \eqref{exputl.equ1}, its
discretization scheme is given by
$$
\mathbbm{V}^\delta(\chi_i, \mathbbm{p}(\chi_i), \mathbbm{x}(\chi_i), 
\vartheta, \mathbbm{g}(\chi_i) ) = 1- \exp \Big( -\gamma
\mathbbm{g}(\chi_i) e^{r (T-\chi_i)} \Big) \mathbbm{H}^\delta(\chi_i,
\mathbbm{p}(\chi_i), \mathbbm{x}(\chi_i),  \vartheta). 
$$

\noi Then the 
discretization scheme \eqref{var.equ4} can be reduced to 
\begin{equation}\label{var.equ5}
\begin{aligned}
& \mathbbm{H}^\delta(\chi_i, \mathbbm{p}(\chi_i), \mathbbm{x}(\chi_i),  
\vartheta) = \min \Big\{ \ F_b(\mathbbm{p}(\chi_i), \mathbbm{x}(\chi_i), 
\xi) \cdot \mathbbm{H}^\delta(
\chi_i, \mathbbm{p}(\chi_i), \mathbbm{x}(\chi_i), \vartheta +\xi), \\ 
& \hspace{5pt} F_s(\mathbbm{p}(\chi_i), \mathbbm{x}(\chi_i), \xi) \cdot
\mathbbm{H}^\delta(\chi_i, \mathbbm{p}(\chi_i), \mathbbm{x}(\chi_i),  
\vartheta -\xi),  \ E \big\{ \mathbbm{H}^\delta \big(\chi_{i+1},  
\mathbbm{p}(\chi_{i+1}),  \mathbbm{x}(\chi_{i+1}),
\vartheta \big) \big\} \ \Big\}.
\end{aligned}
\end{equation}

\noi where
$$
F_b(\mathbbm{p}(\chi_i), \mathbbm{x}(\chi_i), \xi) = \exp \big\{ \gamma \xi
A_- (\mathbbm{p}(\chi_i), \mathbbm{x}(\chi_i))
e^{r(T-\chi_i)} \big \},
$$
$$
F_s(\mathbbm{p}(\chi_i), \mathbbm{x}(\chi_i), \xi) = \exp \big\{ -\gamma \xi
A_+ (\mathbbm{p}(\chi_i), \mathbbm{x}(\chi_i))
e^{r(T-\chi_i)} \big \}.
$$

\section{Simulation studies}

\subsection{Buy and sell regions}

We use the numerical algorithm proposed in Section \ref{num.alg} to 
studies the buy and sell boundaries of the pairs trading strategy. 
Our study focuses on two aspects of the problem. The first is 
the property of buy and sell boundaries (or no transaction regions)
for a given set of model parameters, and the other is the impact of
different model parameters on the shape of buy and sell boundaries. 
Without loss of the generality, we assume the time horizon $T=1$ and 
$p(0) = 1$ in all our simulation studies. 

\begin{figure}
\begin{center}
\includegraphics[height=14cm]{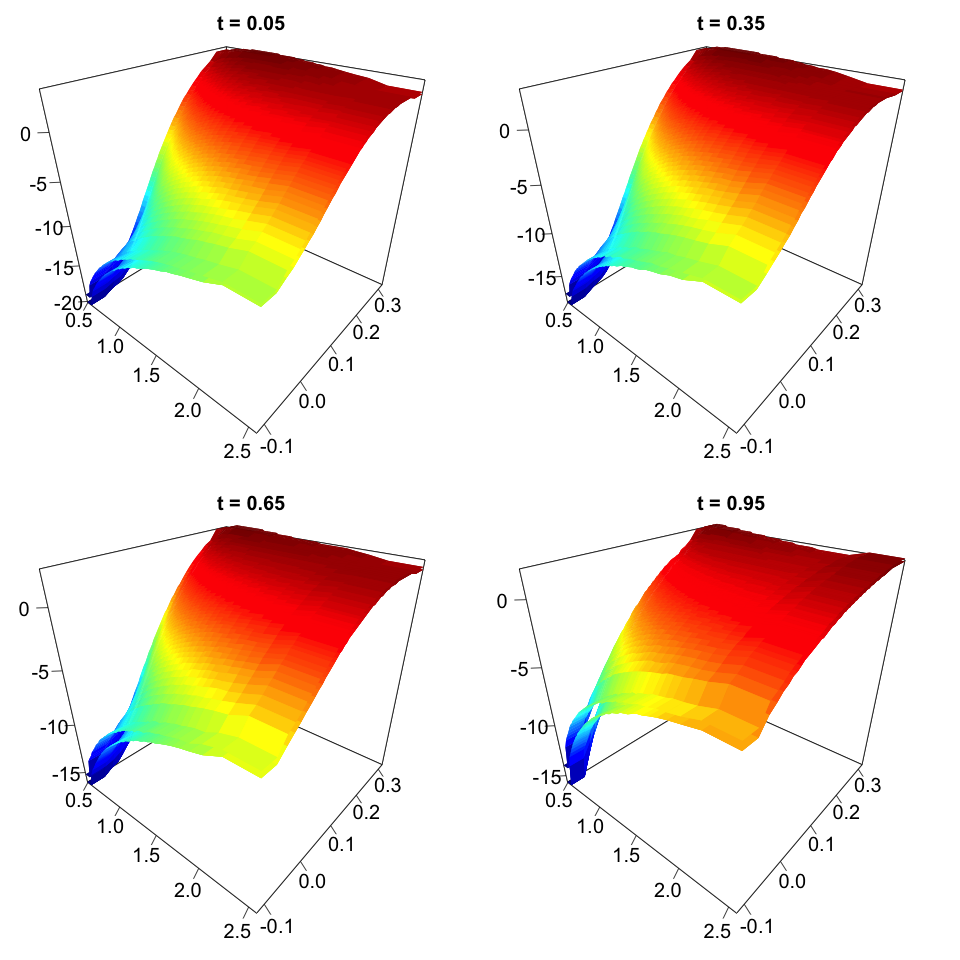}
\caption{Buy and sell boundaries of the baseline scenario (S1) at 
different times.}\label{fig1.bound01}
\end{center}
\end{figure}

\begin{figure}
\begin{center}
\includegraphics[height=14cm]{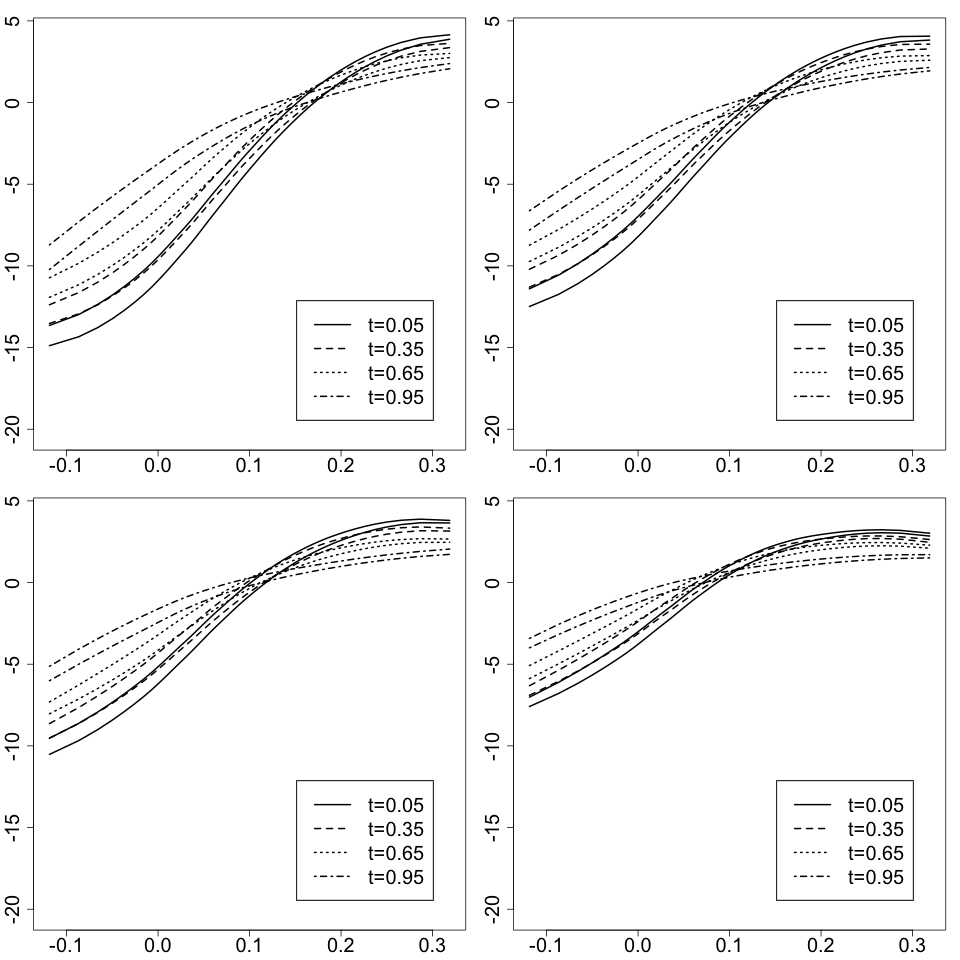}
\caption{Buy and sell boundaries of at prices $P_t=$ 0.845 (top left), 
1.095 (top right), 1.400 (bottom left), and 2.108 (bottom right) 
and different times.}\label{fig2.bound01}
\end{center}
\end{figure}

\begin{figure}
\begin{center}
\includegraphics[height=14cm]{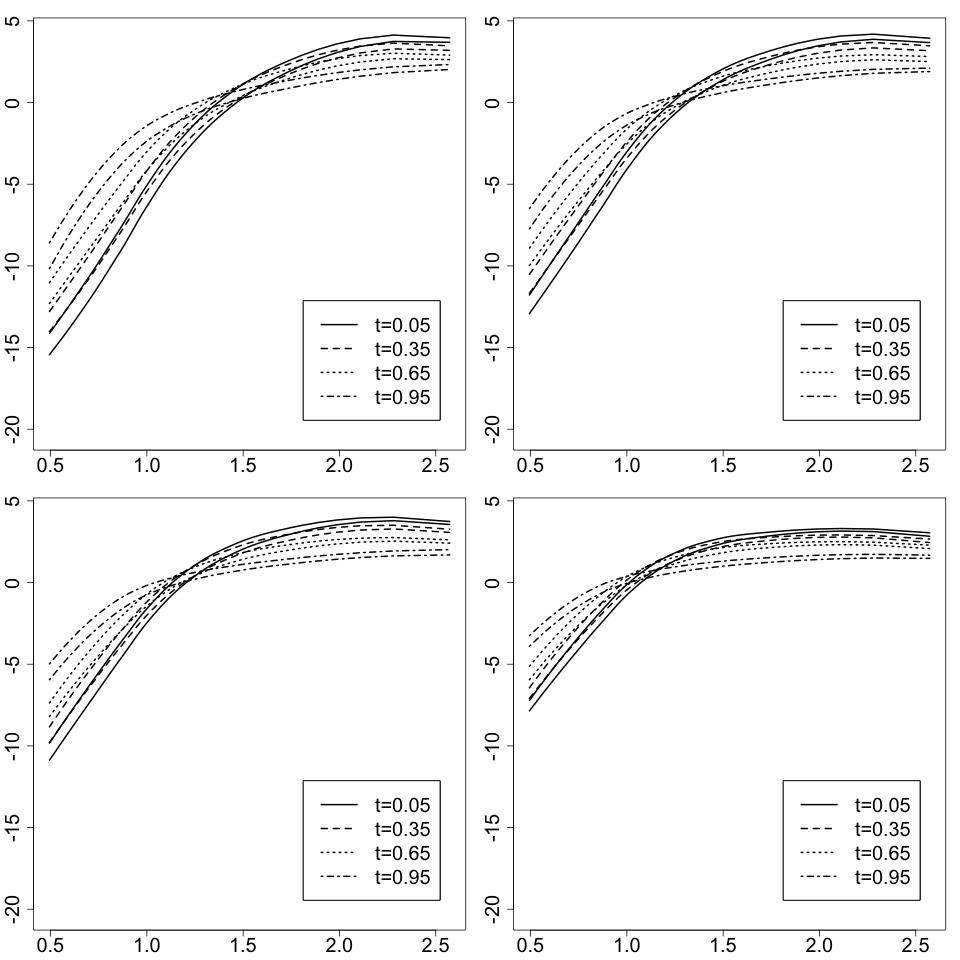}
\caption{Buy and sell boundaries of at spread $X_t=$ 0.023 (top left), 
0.092 (top right), 0.157 (bottom left), and 0.266 (bottom right)
and different times.}\label{fig3.bound01}
\end{center}
\end{figure}

We first consider a baseline scenario. 
The parameter values in the baseline scenario are 
$\mu=0.2, \sigma=0.4, \theta = 0.1, \kappa=1, \nu=0.15, 
\rho=0.5, r=0.01, \gamma=5$ and $\zeta_p=\zeta_q=\xi_p=\xi_q
= 0.0005$. For convenience, we label the setting of the
baseline parameter values as Scenario 1 or (S1). We  
discretize the state space $(t, p, x, y, g)$ and use the
developed Markov chain approximation to solve the
discretized optimization problem. Figure \ref{fig1.bound01}
shows the buy and sell surfaces of (S1) at time 
$t=0.05, 0.35, 0.65$, and $0.95$. To better read the figure, 
we also show in Figures \ref{fig2.bound01} and \ref{fig3.bound01} 
the buy and sell boundaries of (S1) at prices 
$p=0.845, 1.095, 1.400, 2.108$, and $x = 0.023,
0.092, 0.157, 0.266$, respectively. These points are chosen 
such that they correspond to the 24\%, 48\%, 72\%, and 
96\% quantiles of the distribution of $p(T)$ and asymptotic
distribution of $x(t)$, respectively. We find the following from 
these figures. 
First, at a given time and a given price level, the no transaction region 
becomes narrower when the spread gets larger, and the no transaction 
region moves from the negative  to the positive 
 when the spread turns from the negative to 
the positive. For example, at $t=0.05$ and 
$p(t) = 0.845$, the no transaction region changes from 
$[-9.4, -8.0]$ at $x(t)=0.023$ to $[-4.6, -3.4]$ at 
$x(t) = 0.092$, $[-0.7, 0.2]$ at $x(t) = 0.157$, and
$[3.2, 3.7]$ at $x(t)=0.266$. 
Second, at a given time and a given spread level, the no transaction
region becomes narrower when the price $p(t)$ gets larger, and
the no transaction region moves up
when the price becomes larger. For instance, at $t=0.05$
and $x(t)=0.023$, the no transaction region changes from 
$[-9.4, -8.0]$ at $p(t) = 0.845$ to $[-6.8, -5.6]$ at $p(t)=1.095$,
$[-4.9, -3.9]$ at $p(t) = 1.400$, and $[-2.7, -2.0]$ at $p(t)=2.108$.
Note that, the movement of the no transaction region with respect to
price change but with a fixed spread level is relatively smaller than 
that with respect to spread change but with a fixed price level. Third,
when time ellapses from 0 to 1, the no transaction 
region  moves upward. For instance, at the fixed price-spread level 
$(p(t), x(t)) = (1.095, 0.092)$,
the no transaction intervals at $t=0.05, 0.35, 0.65$ and
$0.95$ are $[-2.6, -1.6]$, $[-2.1, -1.2]$, $[-1.5, -0.7]$,
and $[-0.8, -0.2]$, respectively.

We then discuss the impact of different parameter values on the buy and
sell boundaries (or no transaction regions). Besides the parameter values in (S1), 
we now consider other 18 sets of parameter values, labeled as Scenarios 2-19. 
In each of Scenarios 2-19, all parameters values are same as those
in (S1) except one parameter is changed as the specification; see 
Table \ref{table_para} that summarizes parameter values in all 19 
scenarios. For example, Scenario 2 uses parameter values $\mu=0.1$
and assume all other parameters $\sigma, \theta, \kappa, \nu, 
\rho, r, \gamma$ and $\zeta_p=\zeta_q=\xi_p=\xi_q$ have same
values as those in (S1). We discretize the state space $(t, p, x, y, g)$, 
and use the developed Markov chain approximation to solve the
discretized optimization problem for Scenarios 2-19.

\begin{table}[tb]
\begin{center}
\caption{Parameter values of different scenarios}\label{table_para}
\vspace{3mm}
\begin{tabular}{| l | l | l |}\hline \hline
\multicolumn{3}{| l|}{
(S1) $\mu=0.2, \sigma=0.4, \theta = 0.1, \kappa=1, \nu=0.15, 
\rho=0.5$, } \\ 
\multicolumn{3}{| l|}{ \hspace{20pt}
 $r=0.01, \gamma=5$ and $\zeta_p=\zeta_q=\xi_p=\xi_q
= 0.0005$. } \\ \hline
(S2) $\mu=0.1$ & (S8) $\kappa=0.8$ & (S14) $r=0.005$\\
(S3) $\mu=0.3$ & (S9) $\kappa=1.2$ & (S15) $r=0.03$ \\
(S4) $\sigma=0.2$ & (S10) $\nu=0.1$ & (S16) $\gamma = 3$ \\
(S5) $\sigma=0.6$ & (S11) $\nu=0.2$ & (S17) $\gamma = 8$ \\
(S6) $\theta = -0.05$ & (S12) $\rho= -0.2$ & (S18) $\zeta_p=\zeta_q=\xi_p=\xi_q=0.0001$\\
(S7) $\theta=0.3$ & (S13) $\rho=0.6$ & (S19) $\zeta_p=\zeta_q=\xi_p=\xi_q=0.0010$\\ 
\hline \hline
\end{tabular}
\end{center}
\end{table}

\begin{figure}
\begin{center}
\includegraphics[height=14cm]{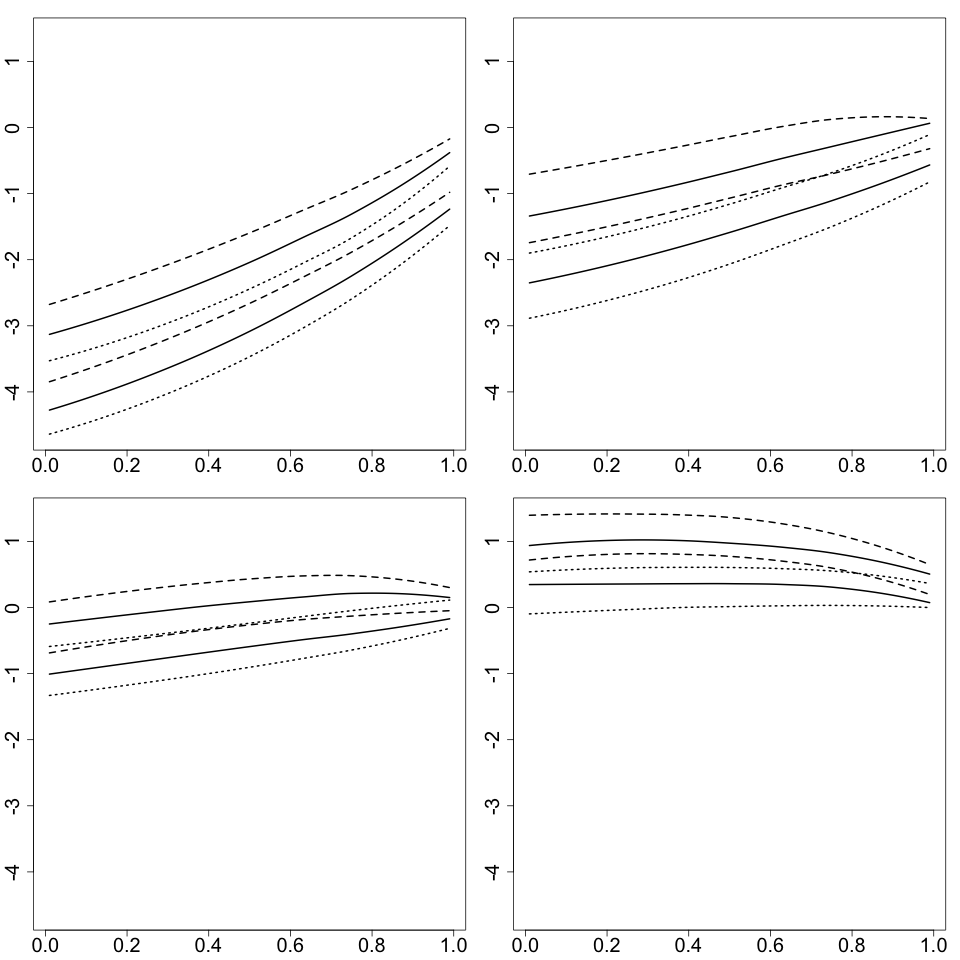}
\caption{Buy and sell boundaries of at fixed prices $(p^{(i)}, x^{(i)})$,
$i=1, 2, 3, 4$ for $\mu=$ 0.1 (dashed), 0.2 (solid), 0.3 (dotted).}\label{fig4.mu}
\end{center}
\end{figure}

\begin{figure}
\begin{center}
\includegraphics[height=14cm]{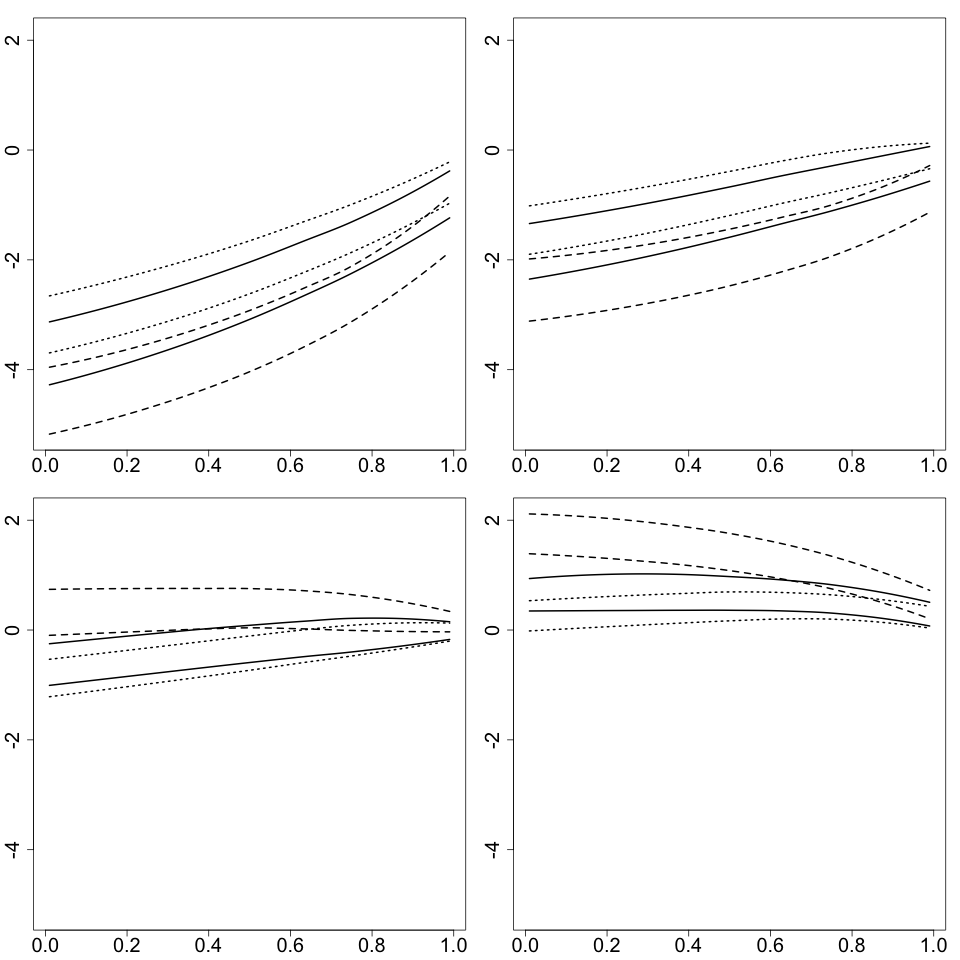}
\caption{Buy and sell boundaries of at fixed price $(p^{(i)}, x^{(i)})$,
$i=1, 2, 3, 4$ 
for $\sigma=$ 0.2 (dashed), 0.4 (solid),  0.6 (dotted).}\label{fig4.sigma}
\end{center}
\end{figure}

\begin{figure}
\begin{center}
\includegraphics[height=14cm]{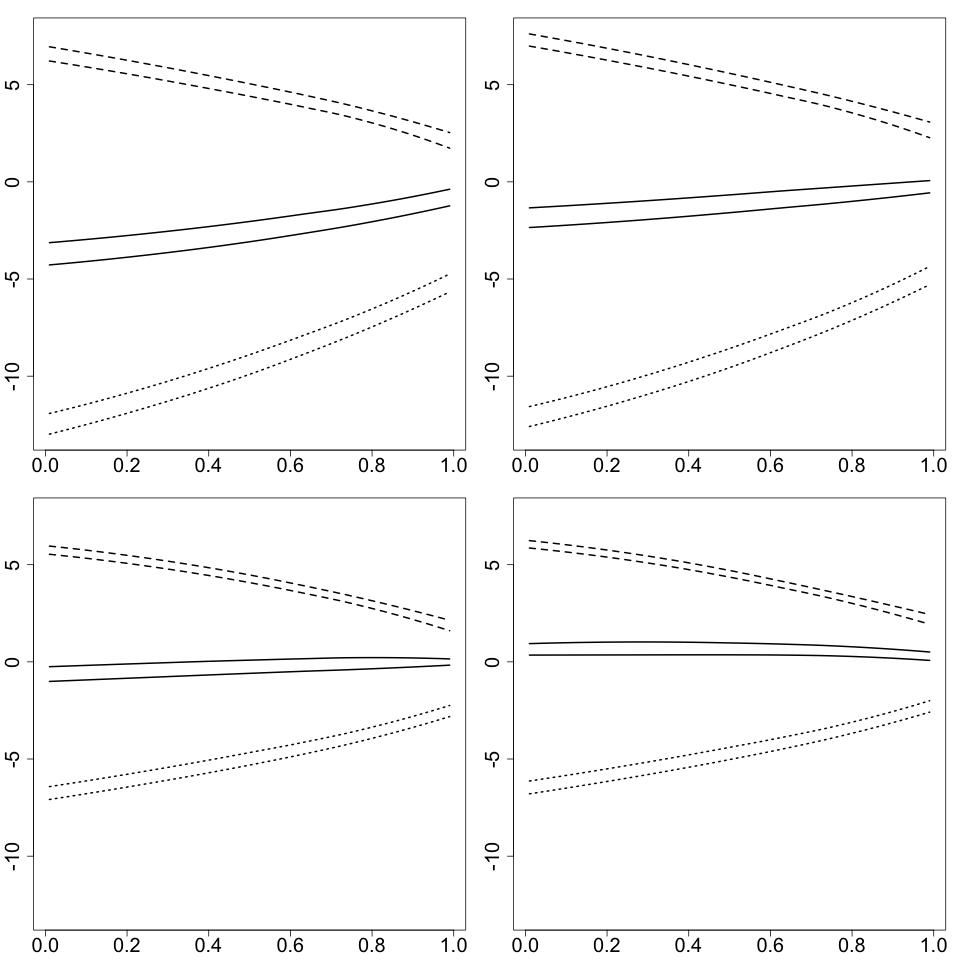}
\caption{Buy and sell boundaries of at fixed price $(p^{(i)}, x^{(i)})$,
$i=1, 2, 3, 4$ 
 for $\theta= -0.05$ (dashed), 0.1 (solid), 0.3 (dotted).}\label{fig4.theta}
\end{center}
\end{figure}

\begin{figure}
\begin{center}
\includegraphics[height=14cm]{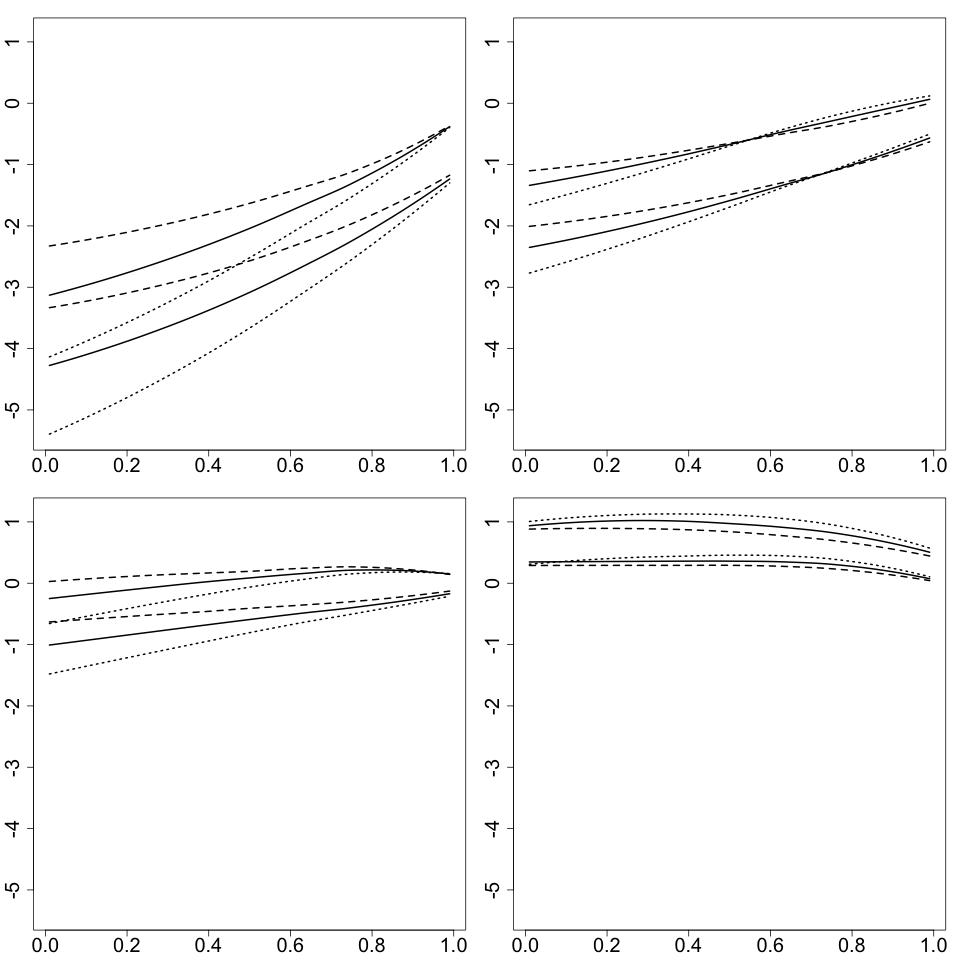}
\caption{Buy and sell boundaries of at fixed price $(p^{(i)}, x^{(i)})$,
$i=1, 2, 3, 4$ 
 for $\kappa=$ 0.8 (dashed), 1 (solid),  and 1.2 (dotted).}\label{fig4.kappa}
\end{center}
\end{figure}

\begin{figure}[t]
\begin{center}
\includegraphics[height=14cm]{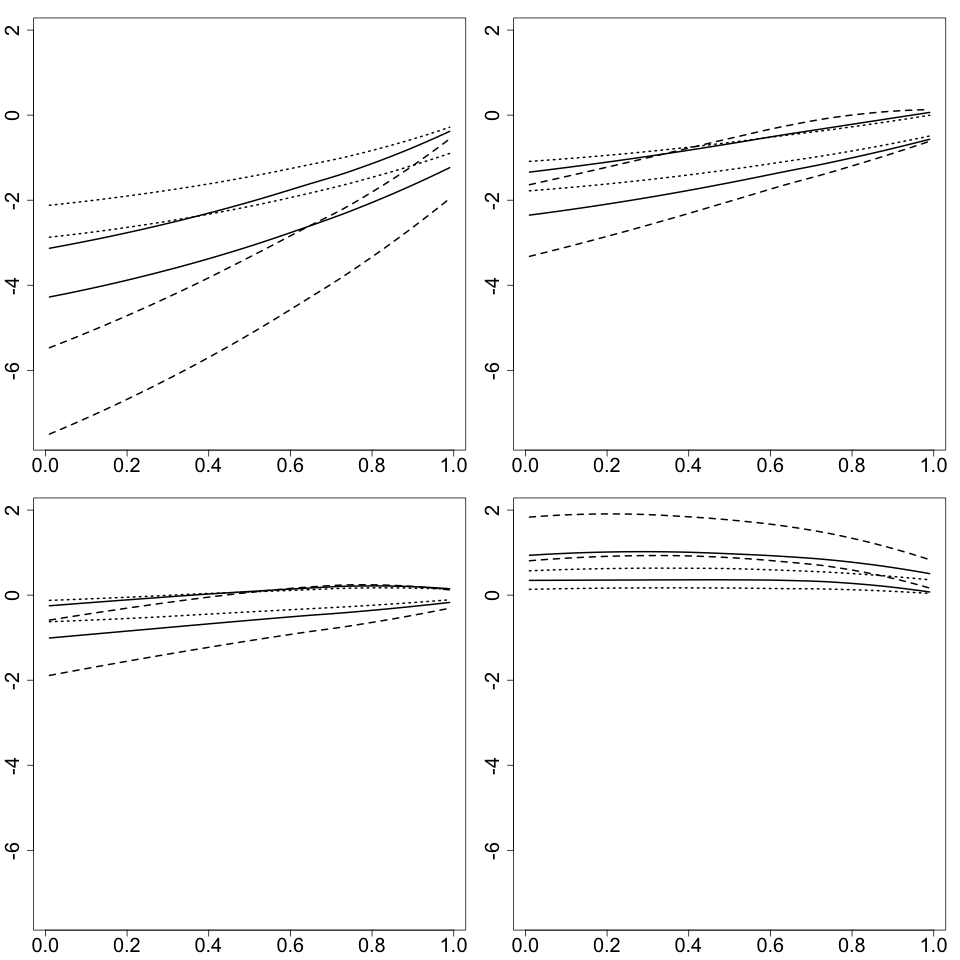}
\caption{Buy and sell boundaries of at fixed price $(p^{(i)}, x^{(i)})$,
$i=1, 2, 3, 4$ 
for $\nu=$ 0.1 (dashed), 0.15 (solid),  and 0.2 (dotted).}\label{fig4.nu}
\end{center}
\end{figure}

\begin{figure}
\begin{center}
\includegraphics[height=14cm]{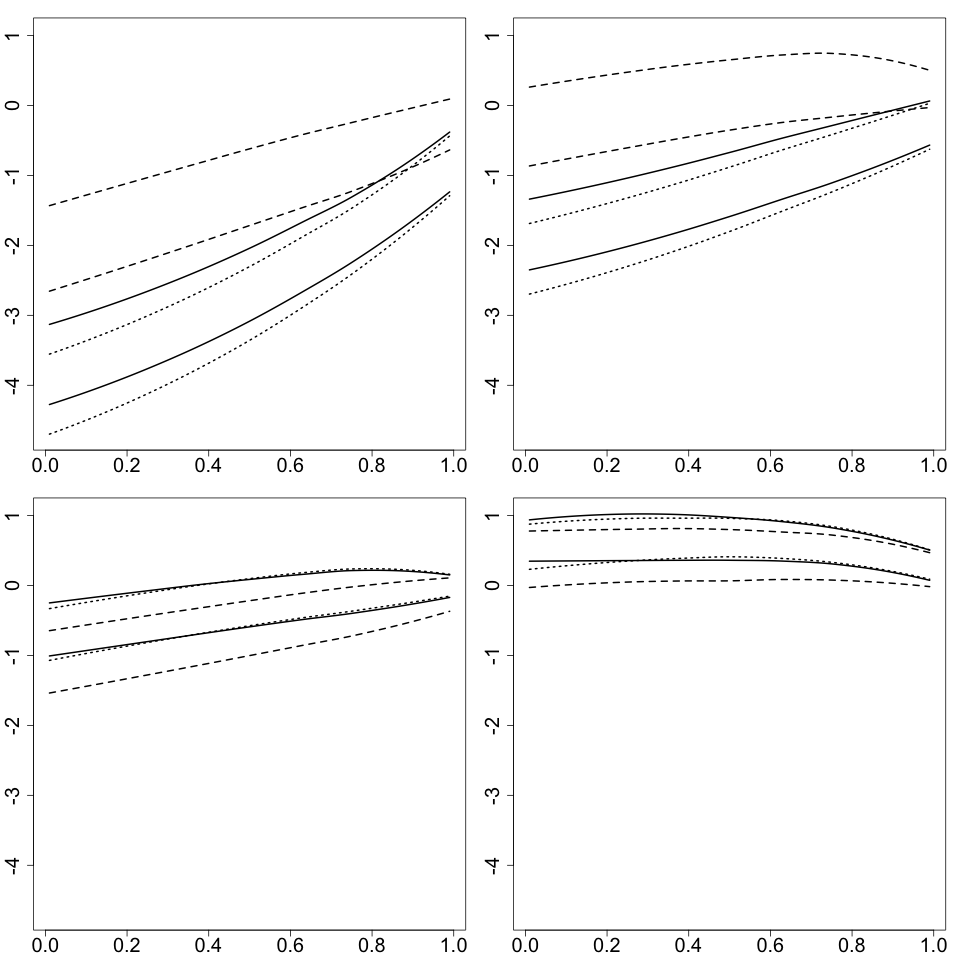}
\caption{Buy and sell boundaries of at fixed price $(p^{(i)}, x^{(i)})$,
$i=1, 2, 3, 4$ 
 for $\rho= -0.2$ (dashed), 0.5 (solid), and 0.6 (dotted).}\label{fig4.rho}
\end{center}
\end{figure}

\begin{figure}
\begin{center}
\includegraphics[height=14cm]{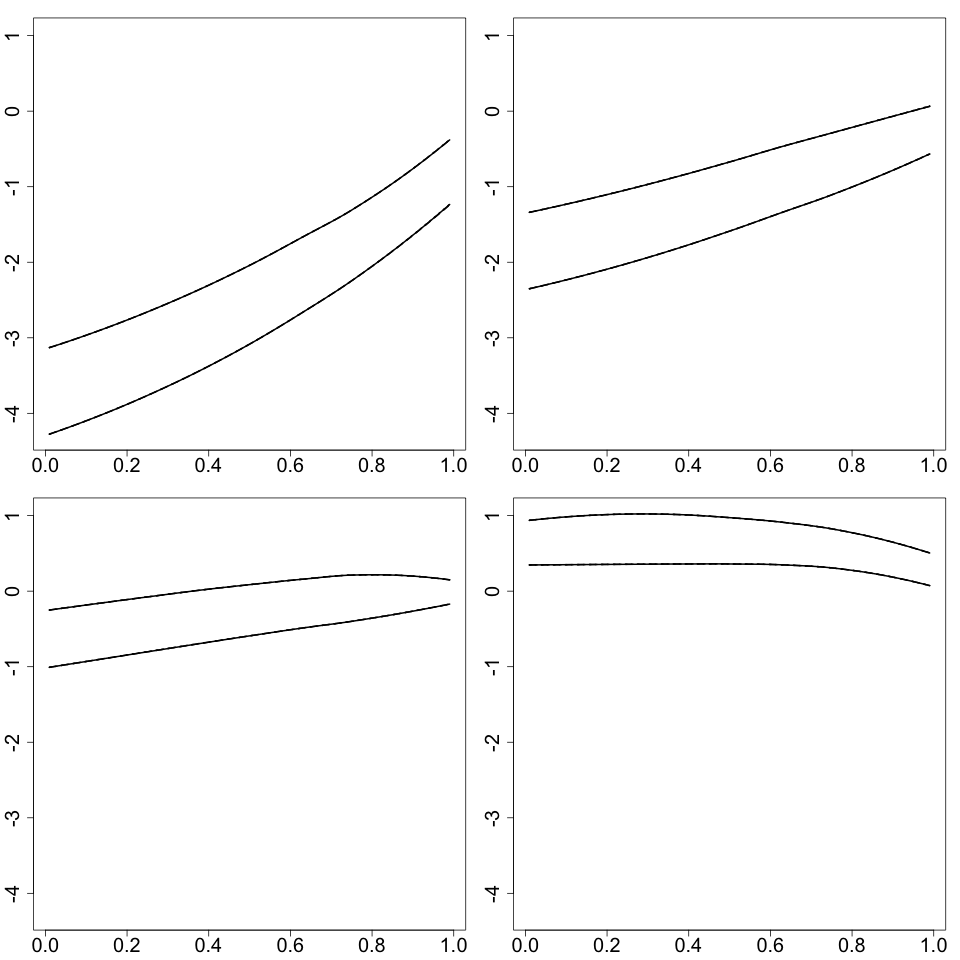}
\caption{Buy and sell boundaries of at fixed price $(p^{(i)}, x^{(i)})$,
$i=1, 2, 3, 4$ 
for $r=$ 0.005 (dashed), 0.01 (solid), and 0.03 (dotted).}\label{fig4.r}
\end{center}
\end{figure}

\begin{figure}
\begin{center}
\includegraphics[height=14cm]{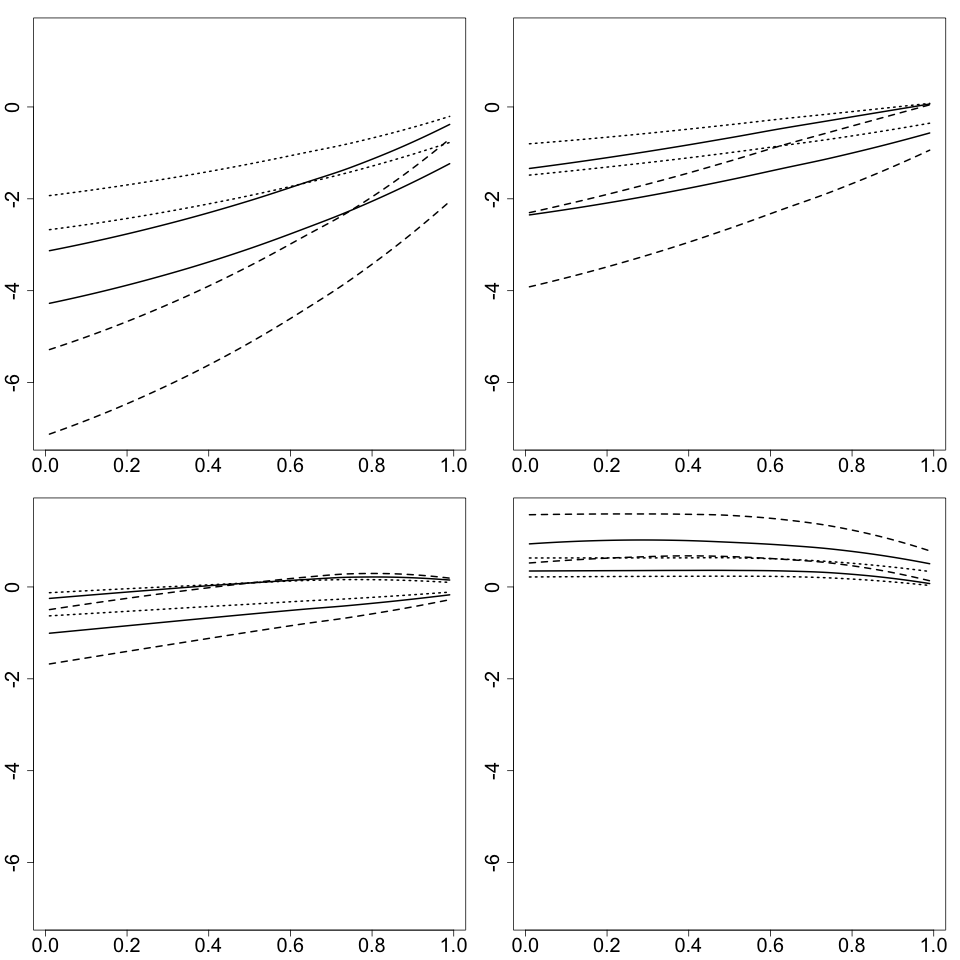}
\caption{Buy and sell boundaries of at fixed price $(p^{(i)}, x^{(i)})$,
$i=1, 2, 3, 4$  for $\gamma =$ 3 (dashed), 5 (solid), and 8 (dotted).}\label{fig4.gamma}
\end{center}
\end{figure}

\begin{figure}
\begin{center}
\includegraphics[height=14cm]{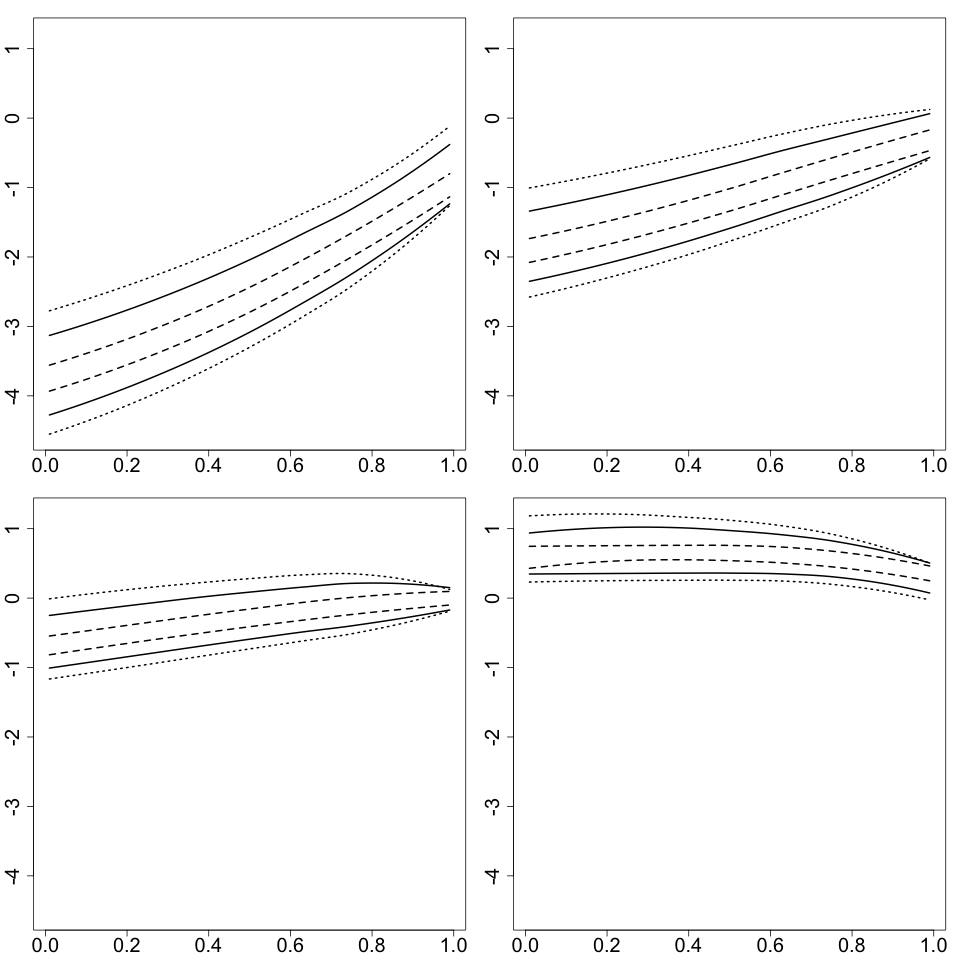}
\caption{Buy and sell boundaries of at fixed price $(p^{(i)}, x^{(i)})$,
$i=1, 2, 3, 4$ for $\zeta_p =\zeta_q=\xi_p=\xi_q =$ 0.0001 (dashed), 
0.0005 (solid),  and 0.0010 (dotted).}\label{fig4.xi}
\end{center}
\end{figure}

To compare the buy and sell boundaries (or no transaction regions)
among different scenarios, we plot the buy and sell boundaries over 
time at four fixed points $(p^{(1)}, x^{(1)}) = (0.9, 0.09)$, 
$(p^{(2)}, x^{(2)}) =(0.9, 0.12)$, $(p^{(3)}, x^{(3)}) =(1.5, 0.09)$, 
and $(p^{(4)}, x^{(4)}) =(1.5, 0.12)$, respectively. Figures \ref{fig4.mu}-\ref{fig4.xi}
demonstrate variations of the buy and sell boundaries over time for 
different values of $\mu$, $\sigma$, $\theta$, $\kappa$, $\nu$,
$\rho$, $r$, $\gamma$, $\zeta_p (=\zeta_q=\xi_p=\xi_q)$, respectively.
In each figure, we plot the buy and sell boundaries for $(p^{(i)}, x^{(i)})$,
$i=1, 2, 3, 4$ on the top left, top right, bottom left, and bottom right, 
respectively, we also use the solid (dashed, dotted) lines to represent the baseline 
value (the smaller value, the larger value) of the parameter under 
comparison.
Figure \ref{fig4.mu} suggests that when $\mu$ increases, the 
buy and sell boundaries move downward at all four points. 
Figure \ref{fig4.sigma} indicates that when $\sigma$ increases, 
the buy and sell boundaries move upward at $(p^{(1)}, x^{(1)})$
and  $(p^{(2)}, x^{(2)})$, but move downward at  $(p^{(3)}, x^{(3)})$
and  $(p^{(4)}, x^{(4)})$. Figure \ref{fig4.theta} shows that, when
$\theta$ increases, the buy and sell boundaries move downward at 
all four points. Figure \ref{fig4.kappa} indicates that, when $\kappa$ 
increases, the buy and sell boundaries move downward, and the
magnitude of such movement is larger at $(p^{(1)}, x^{(1)})$ 
than the other three points. Figure \ref{fig4.nu} shows that,
when $\nu$ increases, the buy and sell boundaries move upward
at $(p^{(i)}, x^{(i)})$, $i=1,2,3$, but move downward at $(p^{(4)}, x^{(4)})$.
Figure \ref{fig4.rho} suggests that, when the correlation $\rho$ 
changes from the negative to the positive, the buy and sell boundaries
move downwards at $(p^{(1)}, x^{(1)})$ and  $(p^{(2)}, x^{(2)})$, 
but move upward at  $(p^{(3)}, x^{(3)})$ and  $(p^{(4)}, x^{(4)})$.
Figure \ref{fig4.r} indicates that variations of interest rate $r$ have little
impact on the buy and sell boundaries. Figure \ref{fig4.gamma} shows
that, when the risk aversion parameter $\gamma$ increases, the buy 
and sell boundaries move upward at $(p^{(i)}, x^{(i)})$, $i=1,2,3$, 
but move downward at $(p^{(4)}, x^{(4)})$. Figure \ref{fig4.xi}
suggests that, when the transaction cost increases, the center of
the no transaction region seems not change, but the region gets 
wider. 

\subsection{Performance of the strategy}

We also perform simulation studies to investigate the performance of
the optimal trading strategy. For comparision purpose,  
we also consider a benchmark strategy which is
analogous to the relative-value arbitrage strategy used in 
\cite{GatevEtAl2006} and based on standard deviation of the spread.
Specifically, the strategy opens a position 
when the spread exceeds twice of the standard deviation of
the spread process, and closes the position when either 
price converges or the maturity is reached. As the benchmark 
strategy doesn't specify the number of shares of stocks that should
be bought or sold, we assume that the number of shares of stocks
traded each time is one. 

We simulate the price process $p_t$ and the spread process $x_t$
to compare the performance of the benchmark strategy and our
strategy in scenarios (S1)-(S19). Assume that $T=1$, and we 
discretize the time interval $(0, 1]$ as $\{ 0.01, 0.02, \dots, 0.99, 1\}$, 
so that we have 100 trading periods. For each scenario, we simulate
1000 paths of $\{ (p_t, x_t) | t=0, 0.01, \dots, 0.99, 1, p_0=1\}$, and 
for each simulated path $(p_t, x_t)$, we implement the benchmark 
strategy and the optimal strategy at $t=0.01, 0.02, \dots, 0.99$ and 
close the position at $T=1$. Let $i=b$, $o$ represent
the benchmark and the optimal strategies, respectively. For each
realized trading strategies, denote $N^{(i)}$ as the number of 
trades (i.e., buy and sell) among the 100 trading periods and
$PL^{(i)}= -C^{(i)}(L_p, M_p; 0,1)$ the total profit made during 
the trading process. Note that the benchmark strategy trades only 
one share of stock each time while the number of shares of stocks in the
optimal strategy are ``optimally" chosen based on the buy and sell 
regions, we define $PS^{(i)}$ as the the average profit (or loss)
generated from the maximum number of shares of stocks during the
trading process. That is, 
$PS^{(i)} := -C^{(i)}(L_p, M_p; 0,1)/ \max_{t} |Y_t^{(i)}|$,
where $Y_t^{(i)}$ is the number of shares of stock $P$ at 
$t=0.01, 0.02, \dots, 0.99$.

Table \ref{table_comp} summarizes 
the mean and standard error of $N^{(i)}$, $PL^{(i)}$, and
$PS^{(i)}$ ($i=o, b$) for 1000 paths
in each scenario. We note that the total numbers of trades $N^{(o)}$
in the optimal strategy range from 45.736 to 55.821 for 
(S1)-(S17), and increases (or decreases) significantly when the 
transaction costs decreases (or increases) in (S18) and (S19). 
In comparison to this, the total numbers of trades $N^{(b)}$
in the benchmark strategy are much smaller, essentially, 
between 1 and 2. This suggests the benchmark strategy is 
much more conservative than the optimal strategy. For the
realized profit over the trading period, $PL^{(o)}$ is much 
larger than $PL^{(b)}$ as the optimal strategy can 
choose to buy or sell the ``optimal" number of shares of stock 
pairs, while the benchmark strategy only buy or sell one share of 
stock pair. $PS^{(o)}$ and $PS^{(b)}$ remove the impact of 
number of shares of traded stocks, and provide the average 
earning per traded stock, and we notice that $PS^{(o)}$ still
significantly higher than $PS^{(b)}$.

\begin{table}[tb]
{\footnotesize
\begin{center}
\caption{Performance of strategies}\label{table_comp}
\begin{tabular}{c|ccc|ccc}\hline \hline
& $N^{(o)}$ & $PL^{(o)}$ & $PS^{(o)}$ &  
$N^{(b)}$ & $PL^{(b)}$ & $PS^{(b)}$ \\ \hline
(S1) & 52.289 (.247) & 0.349 (.019) & 0.048 (.004) & 1.094 (.084) & 0.005 (.002) & 0.005 (.002)\\
(S2) & 53.218 (.241) & 0.389 (.020) & 0.051 (.004) & 1.094 (.084) & 0.006 (.002) & 0.006 (.002)\\
(S3) & 51.348 (.253) & 0.318 (.019) & 0.046 (.004) & 1.094 (.084) & 0.004 (.002) & 0.004 (.002)\\
(S4) & 52.999 (.208) & 0.378 (.019) & 0.054 (.003) & 1.094 (.084) & 0.007 (.002) & 0.007 (.002)\\
(S5) & 51.896 (.275) & 0.326 (.019) & 0.040 (.005) & 1.094 (.084) & 0.003 (.004) & 0.003 (.004)\\
(S6) & 49.299 (.235) & 0.357 (.020) & 0.032 (.003) & 1.094 (.084) & 0.003 (.002) & 0.003 (.002)\\
(S7) & 54.233 (.262) & 0.344 (.019) & 0.064 (.005) & 1.094 (.084) & 0.008 (.003) & 0.008 (.003)\\
(S8) & 55.821 (.304) & 0.359 (.021) & 0.062 (.006) & 1.094 (.084) & 0.005 (.003) & 0.005 (.003)\\
(S9) & 45.736 (.254) & 0.266 (.016) & 0.046 (.003) & 1.094 (.084) & 0.005 (.002) & 0.005 (.002)\\
(S10) & 46.347 (.292) & 0.228 (.016) & 0.042 (.005) & 1.052 (.083) & 0.004 (.002) & 0.004 (.002)\\
(S11) & 57.689 (.212) & 0.489 (.022) & 0.053 (.003) & 1.206 (.084) & 0.007 (.002) & 0.007 (.002)\\
(S12) & 46.774 (.248) & 0.325 (.015) & 0.065 (.003) & 1.140 (.086) & 0.008 (.001) & 0.008 (.001)\\
(S13) & 53.516 (.245) & 0.361 (.020) & 0.045 (.004) & 1.140 (.087) & 0.006 (.002) & 0.006 (.002)\\
(S14) & 54.027 (.232) & 0.579 (.032) & 0.048 (.004) & 1.094 (.084) & 0.005 (.002) & 0.005 (.002)\\
(S15) & 50.031 (.266) & 0.219 (.012) & 0.049 (.004) & 1.094 (.084) & 0.005 (.002) & 0.005 (.002)\\
(S16) & 52.300 (.247) & 0.347 (.019) & 0.048 (.004) & 1.094 (.084) & 0.005 (.002) & 0.005 (.002)\\
(S17) & 52.261 (.247) & 0.357 (.019) & 0.050 (.004) & 1.094 (.084) & 0.006 (.002) & 0.006 (.002)\\
(S18) & 73.801 (.286) & 0.339 (.019) & 0.045 (.004) & 1.094 (.084) & 0.006 (.002) & 0.006 (.002)\\
(S19) & 42.996 (.222) & 0.339 (.019) & 0.049 (.004) & 1.094 (.084) & 0.004 (.002) & 0.004 (.002)\\
 \hline \hline
\end{tabular}
\end{center} }
\end{table}

\section{Real data studies}

We test our model with real market data in this section. We 
present the sample and explain our methodology first, and 
then show the results and discussion.

\begin{figure}
\begin{center}
\includegraphics[height=14cm]{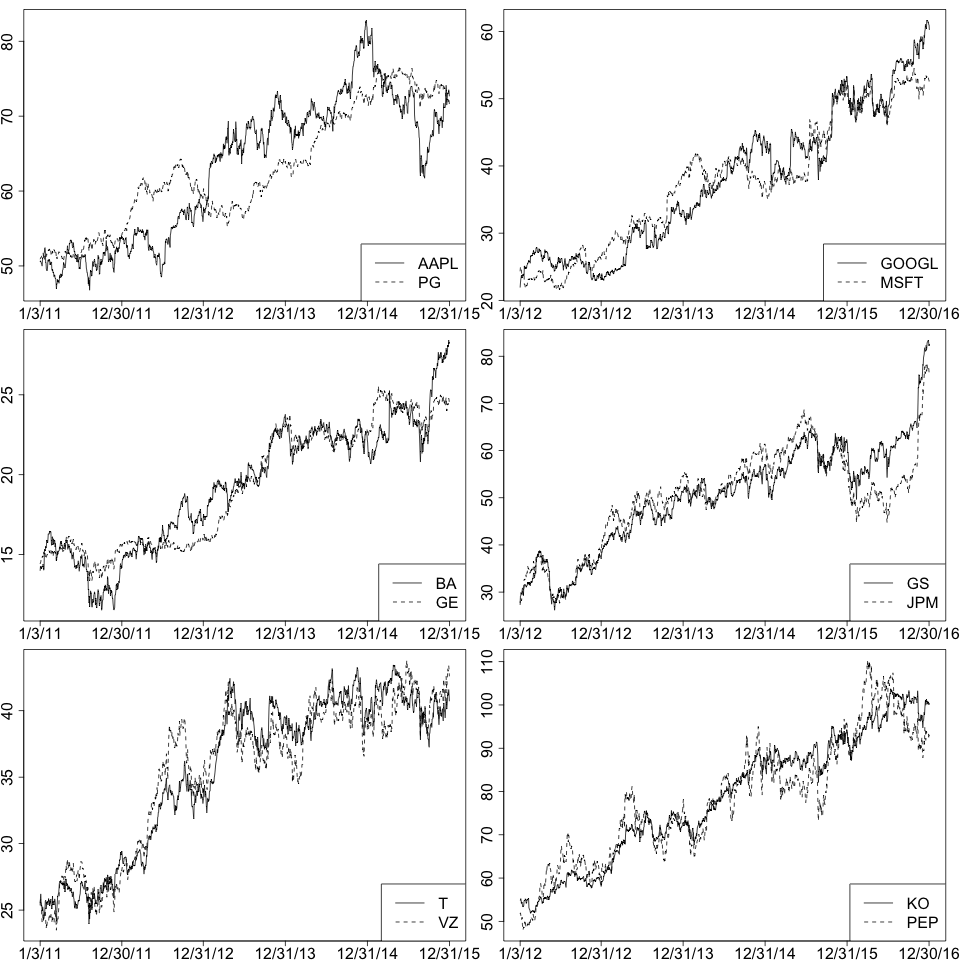}
\caption{Orignal (solid) and transformed (dashed) prices of
six pairs of stocks.}\label{fig5.6stocks}
\end{center}
\end{figure}

A key step of implementing pairs trading strategy is to select 
two stocks for pairs trading. \citep{GatevEtAl2006} illustrate 
how this can be done by using stock price data. An alternative
to this approach is to use fundamentals analysis to select
two stocks that have almost the same risk factor exposures; see
\cite{Vidyamurthy2004}. In this study, we consider a hybrid of
these two approaches. Specifically, we restrict two stocks 
$P$ and $Q$ to belong to the same industry sector. Table
\ref{table_stocks} lists six pairs of stocks selected from 
four different sectors. For each pair of stocks $P$ and $Q$, 
we compute the spread by regressing log price of stock $Q$
on the log price of stock $P$, and the fitted values of the 
regression is considered as the ``transformed" price of $P$.
Figure \ref{fig5.6stocks} shows six pairs of the original prices 
of $Q$ and transformed prices of $P$ over time. 

\begin{table}[tb]
\begin{center}
\caption{Six pairs of stocks selected from different industries.}\label{table_stocks}
\begin{tabular}{l|l|l}\hline \hline
Sector & Stock $Q$ & Stock $P$ \\ \hline
Consumer goods & Apple Inc. (AAPL) &  Procter \& Gamble Co. (PG) \\ 
Consumer goods & Coca-Cola Co (KO) & PepsiCo, Inc. (PEP) \\ 
Technology & Alphabet Inc Class A (GOOGL) & Microsoft Corporation (MSFT) \\
Technology & AT\&T Inc. (T) & Verizon Communications Inc. (VZ)\\
Industrial goods & Boeing Corporation (BA) & General Electric Company (GE)\\
Financial & Goldman Sachs Group Inc. (GS) & JPMorgan Chase \& Co. (JPM)\\\hline \hline
\end{tabular}
\end{center}
\end{table}

We then apply the optimal strategy and the benchmark strategy
in Section 4.2 to test the out-of-the-sample performance. 
Specifically, we use the past three years of the historical
data of each pair to estimate the model parameter, and run 
unit-root test to conclude if the spread $x_t$ is a 
stationary process. If $x_t$ is not stationary, we do not 
implement any strategies. Otherwise, we implement both
the optimal strategy and the benchmark strategy. Note that
the optimal strategy can optimally choose the number of 
shares of stocks in each trade, while we still trade one 
unit of stock in the benchmark strategy. Table
\ref{table_res} shows the number of trades $N^{(i)}$, 
the accumulated profit (in U.S. dollars) at maturity 
$PL^{(i)}$, and the average profit per traded share
$PS^{(i)}$ over two testing periods, 
for $i=o$ (the optimal strategy) and
$i=b$ (the benchmark strategy). Table \ref{table_res} 
suggests that the benchmark strategy is much more 
conservative than the optimal strategy. 
Besides, the average profits per traded share 
$PS^{(o)}$ of the optimal strategy are much larger than that 
of the benchmark strategy except for the stock pair 
$(KO, PEP)$.

\begin{table}[tb]
\begin{center}
\caption{Performance of strategies}\label{table_res}
\begin{tabular}{l|l|rrr|rrr}\hline \hline
Pairs & Year & $N^{(o)}$ & $PL^{(o)}$ & $PS^{(o)}$ &
	$N^{(b)}$ & $PL^{(b)}$ & $PS^{(b)}$ \\ \hline
(AAPL, PG) & 2014 & 58 & 8.56 & 2.173 & 0 & 0 & 0 \\
 & 2015 & 70 & 25.439 & 3.91 & 0 & 0 & 0 \\ \hline
(BA, GE) & 2014 & 97 & 27.866 & 1.292 & 0 & 0 & 0\\
 & 2015 & 165 & 168.543 & 1.982 & 20 & 0.455 & 0.455\\ \hline
(T, VZ) & 2014 & 127 & 77.908 &2.158 &2 &0.603 &0.603\\
 & 2015 & 131 &115.587 &2.883 &0 &0 &0\\ \hline
(GOOGL, MSFT) & 2015 & 103 &94.271 &6.734 &8 &1.623 &1.623\\
& 2016 & 135 & 65.957 & 6.296 & 0 & 0 &0\\ \hline
(GS, JPM) & 2015 & 100 & 7.654 &0.195 &6 &-2.54 &-2.54\\
& 2016 & 200 & 94.542 & 2.375 &8 &-1.66 &-1.66\\ \hline
(KO, PEP) & 2015 & 142 &37.51 &0.675 &22 &10.154 &10.154\\
&2016 & 165 &217.878 &4.059 &4 &5.983 &5.983 \\ \hline \hline
\end{tabular}
\end{center}
\end{table}

\section{Concluding remark}

The problem of optimal pairs trading has been studied by 
many academic researchers and financial practitioners. 
Existing models and methods try to find either the optimal 
shares of stocks by assuming no transaction costs, or the 
optimal timing of trading fixed number of shares of stocks 
with transaction costs. To find optimal pairs trading strategies 
which determine optimally both the trade time and the number 
of shares during the trading process, we investigate an 
optimal pairs trading problem with proportional transaction 
costs. Using an approach that is based on maximization of the 
expected utility of terminal wealth, we transform the problem 
into a singular stochastic control problem and argue that 
the value function of the problem is unique viscosity 
solution of a nonlinear quasi-variational inequality. We 
further show that the viscosity solution is equivalent to a 
free boundary problem for the singular stochastic control 
value function. To solve the singular stochastic control 
problem associated with utility maximization and compute 
the value function and transaction regions, we develop 
a dynamic programming based numerial algorithm to compute 
the solution. In simulation studies, we illustrate the 
numerical algorithm and investigate the impact of model 
parameters on the optimal trading strategies (or the 
transaction regions). We also demonstrate the out-of-sample 
performance of the optimal strategy via an empirical 
study which consists of six pairs of U.S. stocks from 
different industry sectors.

There are several directions in which our approach needs 
further investigation. First, our approach can be easily 
extended for nonexponential utility functions. In such
a case, the optimization problem involves five (instead of 
four) variables, the numerial algorithm in our paper needs 
to be modified to adapt for five variables.
Second, our approach can be extended to solve the optimal cointegration 
trading which involves $n$ stocks with $m$ cointegration relationship. 
Third, many empirical studies suggest that stock price processes can be 
better approximated by incorporating jumps. Using the framework
and algorithms developed in \cite{XingEtAl2017}, the 
method developed here can be extended 
to the case that price processes follow geometric jump-diffusion 
processes. In such a case, the value function of the corresponding variational inequalities
involve integro-differential equations, which can be solved  by 
extending our numerical algorithm.

\section*{Acknowledgement}

The author's research is supported by 
National Science Foundation DMS-1612501.

\bibliographystyle{plainnat}
\bibliography{mybib}

\begin{thebibliography}{14}
\providecommand{\natexlab}[1]{#1}
\providecommand{\url}[1]{\texttt{#1}}
\expandafter\ifx\csname urlstyle\endcsname\relax
  \providecommand{\doi}[1]{doi: #1}\else
  \providecommand{\doi}{doi: \begingroup \urlstyle{rm}\Url}\fi

\bibitem[Ehrman(2006)]{Ehrman2006}
D.~Ehrman.
\newblock \emph{The Handbook of Pairs Trading: Strategies Using Equities,
  Options, and Futures}.
\newblock John Wiley and Sons, New Jersey, 2006.

\bibitem[Elliott et~al.(2005)Elliott, Van~der Hoek, and
  Malcom]{ElliottEtAl2005}
R.~Elliott, J.~Van~der Hoek, and W.~Malcom.
\newblock Pairs trading.
\newblock \emph{Quantitative Finance}, 5:\penalty0 271--276, 2005.

\bibitem[Gatev et~al.(2006)Gatev, Goetzmann, and Rouwenhorst]{GatevEtAl2006}
E.~Gatev, W.~N. Goetzmann, and K.~G. Rouwenhorst.
\newblock Pairs trading: Performance of a relative-value arbitrage rule.
\newblock \emph{The Review of Financial Studies}, 19:\penalty0 797--827, 2006.

\bibitem[Lai and Xing(2008)]{laixing2008}
T.Z. Lai and H.~Xing.
\newblock \emph{Statistical Models and Methods for Financial Markets}.
\newblock Springer, New York, 2008.

\bibitem[Lei and Xu(2015)]{LeiXu2015}
Y.~Lei and J.~Xu.
\newblock Costly arbitrage through pairs trading.
\newblock \emph{Journal of Economic Dynamics and Control}, 56:\penalty0 1--19,
  2015.

\bibitem[Leung and Li(2015)]{LeungLi2015}
T.~Leung and X.~Li.
\newblock Optimal mean reversion trading with transaction costs and stop-loss
  exit.
\newblock \emph{International Journal of Theoretical and Applied Finance},
  18:\penalty0 1550020, 2015.

\bibitem[Mudchanatongsuk et~al.(2008)Mudchanatongsuk, Primbs, and
  Wong]{MudchanatongsukEtAl2008}
S.~Mudchanatongsuk, J.~Primbs, and W.~Wong.
\newblock Optimal pairs trading: A stochastic approach.
\newblock \emph{American Control Conference, IEEE}, 2008.

\bibitem[Ngo and Pham(2016)]{NgoPham2016}
M.~Ngo and H.~Pham.
\newblock Optimal switching for the pairs trading rule: A viscosity solutions
  approach.
\newblock \emph{Journal of Mathematical Analysis and Applications},
  441:\penalty0 403--425, 2016.

\bibitem[Song and Yan(2013)]{SongZhang2013}
Q.~Song and R.~Yan.
\newblock An optimal pairs-trading.
\newblock \emph{Automatica}, 49:\penalty0 3007--3014, 2013.

\bibitem[Tourin and Yan(2013)]{TourinYan2013}
A.~Tourin and R.~Yan.
\newblock Dynamic pairs trading using the stochastic control approach.
\newblock \emph{Journal of Economic Dynamics and Control}, 37:\penalty0
  1972--1981, 2013.

\bibitem[Vidyamurthy(2004)]{Vidyamurthy2004}
G.~Vidyamurthy.
\newblock \emph{Pairs Trading --- Quantitative Methods and Analysis}.
\newblock John Wiley and Sons, New York, 2004.

\bibitem[Whistler(2004)]{Whistler2004}
M.~Whistler.
\newblock \emph{Trading Pairs --- Capturing Profits and Hedging Risk with
  Statistical Arbitrage Strategies}.
\newblock John Wiley and Sons, New York, 2004.

\bibitem[Xing et~al.(2017)Xing, Yu, and Lim]{XingEtAl2017}
H.~Xing, Y.~Yu, and T.~W. Lim.
\newblock European option pricing under geometric levy processes with
  proportional transaction costs.
\newblock \emph{Journal of Computational Finance}, 21:\penalty0 101--127, 2017.

\bibitem[Zhang and Zhang(2008)]{ZhangZhang2008}
H.~Zhang and Q.~Zhang.
\newblock Trading a mean-reverting asset: Buy low and sell high.
\newblock \emph{Automatica}, 44:\penalty0 1511--1518, 2008.

\end{thebibliography}

\section*{Appendix: Proof of Theorems}

{\it Proof of Theorem 1.} In our case, the state ${\bf X}$ is $(s, {\bf x})$, where 
${\bf x}=(p, x, y, g)$. Let ${\bf X}_0=(s_0, p_0, x_0, y_0, G_0)$, it follows that there
exists an optimal trading strategy, dictated by the pair of processes
$(L_p^*(t), M_p^*(t)$, where ${\bf X}_0^*(t) = (t, p_0^*(t), x_0^*(t),
y_0^*(t), g_0^*(t))$ is the optimal trajectory, with ${\bf X}_0^*(s_0)
= {\bf X}_0$. 

(i) First, we prove that $V$ is a viscosity subsolution of
\eqref{var.equ} on $ [0,T] \times \mathbbm{R}^+ \times \mathbbm{R} \times 
\mathbbm{R} \times \mathbbm{R})$. For this, we must show that, for
all smooth functions $\phi({\bf X})$, such athat $V({\bf X})-\phi({\bf
  X})$ has a local maximum at ${\bf X}_0$, the following inequaility
holds: 
\begin{equation}\label{subsol.cond}
\min \Big\{  -{\cal B} \phi({\bf X}_0), {\cal S}\phi({\bf X}_0), 
-{\cal L} \phi({\bf X}_0) \Big\} \le 0.
\end{equation}

Without loss of generality, we assume that $V({\bf X}_0) = \phi({\bf
  X}_0)$ and $V \le \phi$ on $ [0,T] \times \mathbbm{R}^+ \times \mathbbm{R} \times 
\mathbbm{R} \times \mathbbm{R}$. We argue by contradiction: if the
arguments inside the operator of \eqref{subsol.cond} satisfy
$-{\cal B} \phi({\bf X}_0) >0$ and ${\cal S}\phi({\bf X}_0)>0$,
then there exists $\theta>0$, such that $- {\cal L} \phi({\bf
  X}_0) > \theta$. From the fact that $\phi$ is smooth, the above
inequalities become $-{\cal B} \phi({\bf X}) >0$, ${\cal S}\phi({\bf
  X})>0$, and $- {\cal L} \phi({\bf X}) > \theta$,
where ${\bf X}=(t, p, x, y, g) \in \mathscr{B}({\bf X}_0)$, a
neighborhood of ${\bf X}_0$.  In Lemma 1, it is shown that ${\bf
  X}_0^*(t)$ has no jumps, P-a.s., at ${\bf X}_0 ={\bf
  X}_0^*(s_0)$. Hence, $\tau(\omega)$, defined by 
\begin{equation*}
\tau(\omega) = \inf \{ t \in (s_0, T] : {\bf
  X}_0^*(t) \notin \mathscr{B}({\bf X}_0) \}, 
\end{equation*}

\noi is positive P-a.s., and therefore the integral along ${\bf X}_0^*(t)$
\begin{equation}\label{int.equ1}
\begin{aligned}
-\theta \int_{s_0}^{\tau} dt > &  E \int_{s_0}^{\tau} {\cal B}\phi( {\bf X}_0^*(t)) dL^*(t) 
- E \int_{s_0}^{\tau} {\cal S} \phi( {\bf X}_0^*(t)) dM^*(t) + E
\int_{s_0}^{\tau} {\cal L} \phi ({\bf X}_0^*(t))dt \\   
= & E\{ I_1 \} - E\{ I_2 \} + E\{ I_3 \},
\end{aligned}
\end{equation}

\noi where $(L^*(t), M^*(t))$ is the optimal trading strategy at ${\bf
  X}_0$. Applying It$\hat{o}$'s formula to $\phi({\bf X})$, where the
state dynamics are given by (1)-(6), we get
\begin{equation}\label{int.equ2}
E \{ \phi( {\bf X}_0^*(\tau)) \} = \phi({\bf X}_0) + E\{ I_1 \} - E\{
I_2 \} + E\{ I_3 \}.
\end{equation}

\noi Since $V({\bf X}) \le \phi({\bf X})$, for all ${\bf X} \in {\cal
  B}({\bf X}_0)$, and $V({\bf X}_0) = \phi({\bf X}_0)$,
\eqref{int.equ1} and \eqref{int.equ2} yield 
$$
E \{ V( {\bf X}_0^*(\tau)) \} \le V({\bf X}_0) + E\{ I_1 \} - E\{ I_2
\} + E\{ I_3 \} < V({\bf X}_0) -\theta \int_{s_0}^{\tau} dt,
$$

\noi which violates the dynamic programming principle, together with
the optimality of $(L^*(t), M^*(t))$. Therefore, at least one of the
arguments inside the minimum operator of \eqref{subsol.cond} is
nonpositive, and hence the value function is a viscosity subsolution
of \eqref{var.equ}.

(ii) In the second part of the proof, we show that $V$ is a viscosity
supersolution of \eqref{var.equ}. For this, we must show that, for
all smooth functions $\phi({\bf X})$, such that $V({\bf X})-\phi({\bf
  X})$ has a local minimum at ${\bf X}_0$, the following inequaility
holds: 
\begin{equation}\label{supersol.cond}
\begin{aligned}
\min \Big\{  - {\cal B} \phi({\bf X}_0), {\cal S}\phi({\bf X}_0),
-{\cal L}\phi ({\bf X}_0) \Big\} \ge 0, 
\end{aligned}
\end{equation}

\noi where, without loss of generality, $V({\bf X}_0)= \phi({\bf
  X}_0)$ and $V({\bf X}) \ge \phi ({\bf X}) $ on $ [0,T] \times
\mathbbm{R}^+ \times \mathbbm{R} \times \mathbbm{R} \times
\mathbbm{R}$. In this case, we prove that each argument of the
minimum operator of \eqref{supersol.cond} is nonnegative. 

Consider the trading strategy $L(t) = L_0>0$, $s_0 \le t \le T$, and
$M(t) = 0$, $s_0 \le t \le T$. By the dynamic pogramming principle, 
\begin{equation*}
V(s_0, p_0, x_0, y_0, g_0) \ge V(s_0, p_0, x_0, y_0+L_0, g - (a_p - 
b_q e^{X_0}) p_0 L_0).
\end{equation*}

\noi This inequality holds for $\phi(s, p, x, y, g)$ as well, and, by
taking the left-hand side to the right-hand side, dividing by $L_0$,
and sending $L_0 \rightarrow 0$, we get ${\cal B}\phi({\bf X}_0) \le
0$. Similary, by using the trading strategy $L(t) = 0, s_0 \le t \le
T$, and $M(t) = M_0 >0$, $s_0\le t \le T$, the second argument inside
the minimum operator is found to be nonnegative. 

Finally, consider the case where no trading is applied. By the dynamic
programming principle
\begin{equation}\label{dyn.equ1}
E \{ V({\bf X}_0^d(t)) \} \le V(s_0, p_0, x_0, y_0, g_0), 
\end{equation}

\noi where ${\bf X}_0^d(t)$ is the state trajectory of starting at
$s_0$, when $M(t)=L(t)=0$, $s_0 \le t \le T$, given by (1)-(6) as
$$
{\bf X}_0^d(t) = (t, p(t), x(t), y_0, g(t))
$$

\noi and ${\bf X}_0^d(t) \in \mathscr{B}({\bf X}_0)$. Therefore, by
applying It$\hat{o}$'s rule on $\phi(s, X, B, y, G)$, inequality
\eqref{dyn.equ1} yields
$$
E \Bigg\{ \int_{s_0}^t {\cal L}\phi ({\bf X}_0^d(\xi)) d \xi
\Bigg\} \le 0,
$$

\noi and, by letting $t\rightarrow s_0$, the third argument inside the
minimumm operator is found to be nonnegative. This complete the
proof. $\hfill \Box$

\medskip

{\it Lemma 1}. Assume that $-{\cal B} \phi({\bf X}_0) >0$, and denote the event
that the optimal trajectory ${\bf X}_0^*(t)$ has a jump of size
$\epsilon$, along the direction $(0, 0, 0, 1, -( a_p - 
b_q e^{x_0} ) p_0)$ by $A(\omega)$. Assume that the state
(after the jump) is $(s_0, p_0, x_0, y_0+\epsilon, -( a_p - 
b_q e^{x_0} )B_0 \epsilon) \in \mathscr{B}({\bf X}_0)$. Then 
\begin{equation}\label{lemma1.equ1}
\Big( {\cal B} \phi({\bf X}_0) \Big) P(A) \ge 0,
\end{equation}

\noi therefore $P(A)=0$. Similarly, if ${\cal S} \phi({\bf X}_0) >0$, 
then the optimal trajectory has no jumps along the direction 
$(0, 0, 0, -1,  ( b_p -a_q e^{x_0} ) p_0)$, P-a.s. at ${\bf x}_0$. 

\medskip

{\it Proof}. By the principle of dynamic programming, 
\begin{equation*}
\begin{aligned}
&V(s_0, p_0, x_0, y_0, g_0) = E \big\{ V (s_0, p_0, x_0, y_0+\epsilon,
-( a_p - b_q e^{x_0} ) B_0 \epsilon) \big\} \\
&= \int_{A(\omega)} V (s_0, p_0, x_0, y_0+\epsilon,
-( a_p - b_q e^{x_0} ) B_0 \epsilon) dP+ \int_{A(\omega)} V(s_0, p_0, x_0, y_0, g_0) dP,
\end{aligned}
\end{equation*}

\noi and therefore
$$
 \int_{A(\omega)} \Big[ \phi (s_0, p_0, x_0, y_0+\epsilon,
-( a_p - b_q e^{x_0} ) B_0 \epsilon)
- \phi (s_0, p_0, x_0, y_0, g_0) \Big] dP\ge 0,
$$

\noi since $V({\bf X}) \le \phi({\bf X})$ for all ${\bf X} \in {\cal
  B}({\bf X}_0)$ and $V({\bf X}_0)  = \phi({\bf X}_0)$. Therefore, 
$$
\lim \sup_{\epsilon \rightarrow 0} \Big\{ \int_{A(\omega)} 
\frac{ \phi(s_0, p_0, x_0, y_0+\epsilon, 
-( a_p - b_q e^{x_0} ) p_0 \epsilon)
- \phi (s_0, p_0, x_0, y_0, g_0)}{ \epsilon} dP \Big\} \ge 0,
$$

\noi and, by Fatou's lemma, 
$$
\int_{A(\omega)} \lim \sup_{\epsilon \rightarrow 0} \Big\{ 
\frac{ \phi(s_0, p_0, x_0, y_0+\epsilon, - ( a_p - b_q e^{x_0} ) B_0 \epsilon)
- \phi (s_0, p_0, x_0, y_0, G_0)}{ \epsilon} \Big\} dP\ge 0,
$$

\noi which implies \eqref{lemma1.equ1}. $\hfill \Box$

\bigskip
\bigskip

{\it Proof of Theorem 3}. Let
\begin{equation*}\label{dis.sol.equ1}
V^\delta (t, p, x, y, g) = \left\{ \begin{array}{ll}
\mathbbm{V}^\delta (\chi, \mathbbm{p}, \mathbbm{x}, y, \mathbbm{g}) &
\mbox{if } t \in [\chi, \chi + \delta), y\in [\nu,
\nu+\kappa \delta), \\ 
Z(\mathbbm{p}, \mathbbm{x}, y, \mathbbm{g}) & \mbox{if } t=T 
\end{array} \right.
\end{equation*}

\noi and 
\begin{equation}\label{dis.sol.equ2}
\underline{V}({\bf X}) = \lim _{{\bf Y} \rightarrow {\bf X}} 
\inf_{\delta \rightarrow 0} \{ \mathbbm{V}^\delta ({\bf Y}) \} \quad
\mbox{and} \quad 
\overline{V}({\bf X}) = \lim _{{\bf Y} \rightarrow {\bf X}} 
\sup_{\delta \rightarrow 0} \{ \mathbbm{V}^\delta ({\bf Y}) \},
\end{equation}

\noi where ${\bf X}=(t, p, x, y, g)$. We will show that
$\underline{V}({\bf X})$ and $\overline{V}({\bf X}) $ are a viscosity
supersolution and a viscosity subsolution of \eqref{var.equ},
respectively. Combining this with the uniquesness of the viscosity
solution of \eqref{var.equ} yields $\underline{V}({\bf X}) \ge
\overline{V}({\bf X})$ on 
$[0, T] \times \mathbbm{R}^+ \times \mathbbm{R} \times \mathbbm{R}
\times \mathbbm{R}$. The opposite inequality is true by the
definition of  $\underline{V}({\bf X})$ and $\overline{V}({\bf X}) $,
and therefore
$$
\underline{V}({\bf X}) =\overline{V}({\bf X}) = V({\bf X}),
$$

\noi which, together with \eqref{dis.sol.equ2}, also implies the local
uniform convergence of $\mathbbm{V}^\delta$ to $V$.

Note that we only prove that $\underline{V}$ is a viscosity supersolution
of \eqref{var.equ}, as the arguments for $\overline{V}$ is
identical. Let ${\bf X}_0$ be a local minimum of $\underline{V}-\phi$
on $[0, T] \times \mathbbm{R}^+ \times \mathbbm{R} \times \mathbbm{R}
\times \mathbbm{R}$, for $\phi \in C^{1,2}( [0, T] \times \mathbbm{R}^+
\times \mathbbm{R} \times \mathbbm{R} \times \mathbbm{R})$. 
Without loss of generality, we may assume that ${\bf X}_0$ is a strict
local minimum, that $\underline{V}({\bf X}_0) = \phi ({\bf X}_0)$, and
that $\phi \le -2 \times \sup_{\delta} \{ ||
\mathbbm{V}^\delta||_{\infty} \}$ outside the vall $\mathscr{B}({\bf
  X}_0, R)$, $R>0$, where $\underline{V}({\bf X}) - \phi({\bf X}) \ge
0$. 

Then there exist sequences $\delta_n \in \mathbbm{R}^+$ and ${\bf
  Y}_n \in [0, T] \times \mathbbm{R}^+ \times \mathbbm{R} \times
\mathbbm{R} \times \mathbbm{R}$, such that
$$
\delta_n \rightarrow 0, {\bf Y}_n \rightarrow {\bf X}_0,
\mathbbm{V}^{\delta_n} ({\bf Y}_n) \rightarrow \underline{V}({\bf
  X}_0), {\bf Y}_n \mbox{ if a global minimum point of }
\mathbbm{V}_j^{\delta_n} - \phi.
$$

\noi Let $h_n = \mathbbm{V}^{\delta_n} - \phi$;  then 
\begin{equation}\label{lemma3.equ1}
h_n \rightarrow 0 \mbox{ and } \mathbbm{V}_j^{\delta_n}({\bf X}) \ge
\phi({\bf X}) + h_n({\bf X}) \quad \mbox{for any } {\bf X} \in 
[0, T] \times \mathbbm{R}^+ \times \mathbbm{R} \times
\mathbbm{R} \times \mathbbm{R}.
\end{equation}

\noi To show that $\underline{V}$ is a viscosity supersolution of
\eqref{var.equ}, it suffices to show that
\begin{equation}\label{conv.equ}
\min \Big\{  - {\cal B} \phi ({\bf X}_0), {\cal S} \phi ({\bf X}_0), -
{\cal L}\phi({\bf X}_0) \Big\} \ge 0. 
\end{equation}

\noi Let $\mathbbm{Y}_n = (s_i, \mathbbm{p}_n, \mathbbm{x}_n, y_n, 
\mathbbm{g}_n)$, where $s_i \in [ \chi_i, \chi_i+\delta_n)$
and $y_{\delta_n} \in [ \vartheta_n, \vartheta_n+\kappa
\delta_n)$. Denote $\bfbbmY_n^{(0)}=(\chi_n, \mathbbm{p}_n, 
\mathbbm{x}_n,  y_n, \mathbbm{g}_n)$,  
$$
\bfbbmY_n^{(1)}=\big(\chi_n, \mathbbm{p}_n,  \mathbbm{x}_n, 
\vartheta_n + \kappa \delta_n, \mathbbm{g}_n - (a_p -
b_q e^{\mathbbm{x}_n} ) \mathbbm{p}_n\kappa \delta_n
\big),
$$ 
$$\bfbbmY_n^{(2)}=\big(\chi_n, \mathbbm{p}_n,  \mathbbm{x}_n,
\vartheta_n-\kappa \delta_n,  \mathbbm{g}_n + (
b_p -a_q  e^{\mathbbm{x}_n} )\mathbbm{p}_n 
\kappa \delta_n \big).
$$ 

\noi Then 
$$
\mathbbm{V}^{\delta_n} ( \bfbbmY_n^{(0)}) = \max \Big\{ \
\mathbbm{V}^{\delta_n} 
( \bfbbmY_n^{(1)}), \mathbbm{V}^{\delta_n}
( \bfbbmY_n^{(2)}), E \big\{ \mathbbm{V}^{\delta_n}
( \bfbbmY_{n+1}^{(0)}) \big \} \ \Big \}.
$$

\noi Now we look at the following three cases.

{\it Case 1}. It holds that $\mathbbm{V}^{\delta_n} ( \bfbbmY_n^{(0)}) = 
\mathbbm{V}^{\delta_n} ( \bfbbmY_n^{(1)}).$
Then \eqref{lemma3.equ1} implies that 
$$
\mathbbm{V}^{\delta_n} ( \bfbbmY_n^{(0)}) \ge \phi ( \bfbbmY_n^{(1)}) 
+ \mathbbm{V}^{\delta_n} ( \bfbbmY_n^{(0)}) - \phi ( \bfbbmY_n^{(0)}),
$$

\noi and therefore
$$
\begin{aligned}
0 & \ge \lim \inf_n \Big\{ \frac{  \phi ( \bfbbmY_n^{(1)}) - 
 \phi ( \bfbbmY_n^{(0)}) }{ \delta_n }\Big\} \ge \lim
\inf_{\delta \rightarrow 0} \Big\{ \frac{ \phi (\bfbbmY_0^{(1)}) 
- \phi ( \bfbbmY_0^{(0)}) }{ \delta }\Big\} \\
& = \frac{\partial \phi({\bf X}_0)}{\partial y} - ( a_p - 
 e^{x_0(t)} ) p_0(t) \frac{\partial \phi({\bf  x}_0)}{\partial g}. 
\end{aligned}
$$

{\it Case 2}. It holds that 
$\mathbbm{V}^{\delta_n} ( \bfbbmY_n^{(0)}) = 
\mathbbm{V}^{\delta_n} ( \bfbbmY_n^{(2)})$. Arguing similarly to case
1, we get  
$$
0 \ge - \Big( \frac{\partial \phi({\bf X}_0)}{\partial y} - ( b_p -
a_q e^{x_0(t)} ) p_0(t) \frac{\partial \phi({\bf
    X}_0)}{\partial g} \Big). 
$$

{\it Case 3}. It holds that $\mathbbm{V}^{\delta_n} ( \bfbbmY_n^{(0)})
= E \big\{ \mathbbm{V}^{\delta_n} ( \bfbbmY_{n+1}^{(0)}) \big \}$. Then
\eqref{lemma3.equ1} implies that  
$$
\mathbbm{V}^{\delta_n} ( \bfbbmY_n^{(0)}) \ge E \big\{ \phi (
\bfbbmY_{n+1}^{(0)}) \big\} + \mathbbm{V}^{\delta_n} ( \bfbbmY_n^{(0)}) -
\phi (\bfbbmY_{n+1}^{(0)}),
$$

\noi and therefore
$$
0  \ge \lim \inf_n \Big\{ \frac{  \phi ( \bfbbmY_{n+1}^{(0)}) - 
 \phi ( \bfbbmY_n^{(0)}) }{ \delta_n }\Big\} \ge \lim
\inf_{ \delta \rightarrow 0} \Big\{ \frac{ \phi (\bfbbmY_1^{(0)}) 
- \phi ( \bfbbmY_0^{(0)}) }{ \delta }\Big\} 
= {\cal L}\phi({\bf X}_0).
$$

\noi Combining the results in cases 1-3 yields \eqref{conv.equ}, and
the proof is complete. $\hfill \Box$

\end{document}